\documentclass[prb,showpacs,twocolumn,amsmath,amssymb,floatfix,superscriptaddress]{revtex4-2}
\usepackage{graphicx} 
\usepackage{bm}
\usepackage{bbm}
\usepackage{color}
\usepackage{amsmath}
\usepackage{amssymb}
\usepackage{color}
\usepackage{footnote}
\usepackage{comment}
\usepackage{hyperref}
\usepackage{subfigure}
\usepackage{appendix}
\usepackage[applemac]{inputenc}

\begin{document}
\title{Non-Hermitian multiterminal phase-biased Josephson junctions}
\author{Jorge Cayao}
\affiliation{Department of Physics and Astronomy, Uppsala University, Box 516, S-751 20 Uppsala, Sweden}
\author{Masatoshi Sato}
\affiliation{Center for Gravitational Physics and Quantum Information, Yukawa Institute for Theoretical Physics, Kyoto University, Kyoto 606-8502, Japan}

\date{\today} 
\begin{abstract}
We study non-Hermitian Josephson junctions formed by multiple superconductors and discover the emergence of exceptional points entirely determined by the interplay of the distinct superconducting phases and non-Hermiticity due to normal reservoirs.  In particular, in Josephson junctions with three and four superconductors, we find stable lines and surfaces of exceptional points protected by non-Hermitian topology and highly tuneable by the superconducting phases. We also discover that, in Josephson junctions formed by laterally coupled superconductors, exceptional points can result from hybridized Andreev bound states and lead to the enhancement of supercurrents controlled by dissipation. Our work unveils the potential of multiterminal Josephson junctions for realizing higher dimensional topological non-Hermitian superconducting phenomena.
\end{abstract}
\maketitle

\section{Introduction}
The advent of non-Hermitian (NH) topology has spurred great activity in distinct areas of physics due to its potential for designing states of matter with no Hermitian counterpart \cite{el2018non,ozdemir2019parity,doi:10.1080/00018732.2021.1876991}. The intriguing NH topology stems from spectral degeneracies known as exceptional points (EPs), where eigenvalues and eigenvectors coalesce, giving rise to the concept of point gaps in the complex energy plane as a peculiar effect absent in Hermitian setups \cite{PhysRevX.8.031079,PhysRevX.9.041015,OS23}. While in one-dimension EPs appear at single points, they become exceptional lines and surfaces in higher dimensions, originating surprisingly stable topological phases that otherwise do not exist \cite{wiersig2020review,parto2020non,RevModPhys.93.015005,doi:10.7566/JPSCP.30.011098}. 

It is now well-understood that dissipation is the mechanism leading to NH  physics \cite{doi:10.1080/00018732.2021.1876991,Meden_2023}. A notable realistic physical scenario for dissipative effects is coupling closed systems to normal reservoirs, which naturally appears in quantum transport \cite{datta1997electronic}. These ideas have recently been explored in superconducting systems, \cite{pikulin2012topological,PhysRevB.87.235421,Ioselevich_2013,JorgeEPs,avila2019non,PhysRevB.105.094502,PhysRevB.107.035408,PhysRevB.107.104515,PhysRevResearch.4.L022018,PhysRevB.105.155418,cayao2024nonhermitian,PhysRevB.109.L161404}, with intriguing results reporting  Andreev EPs and supercurrents due to the interplay of non-Hermiticity and the Josephson effect in junctions formed by two superconductors having a finite phase difference \cite{cayao2023non,li2023anomalous}; see also \cite{shen2024nonhermitian,beenakker2024josephson,pino2024thermodynamics}.  Although these studies reveal the role of a single superconducting phase difference in NH JJs,  the impact of non-Hermiticity on JJs formed by multiple superconductors with distinct phases remains unknown.  Having  JJs with multiple phases is not only a physical curiosity. Understanding the phases as bandstructure quasimomenta, in the Hermitian regime it has already enabled the realization of higher dimensional states \cite{PhysRevB.92.155437,riwar2016multi,peralta2022multi}, Andreev molecules \cite{Su_2017,10.21468/SciPostPhysCore.2.2.009,PhysRevB.101.174506,matsuo2023phase2,Coraiola_2023,PRXQuantum.5.020340}, and non-local Josephson transport \cite{Pillet_2019,PhysRevX.10.031051,Matsuo_2022,matsuo2023phase,Haxell_2023,PhysRevB.109.205406,10.21468/SciPostPhys.17.2.037}.   It is therefore natural to wonder about the role of non-Hermiticity on multiterminal JJs for realizing higher dimensional  NH topological phases.
      \begin{figure}[!t]
\centering
	\includegraphics[width=0.49\textwidth]{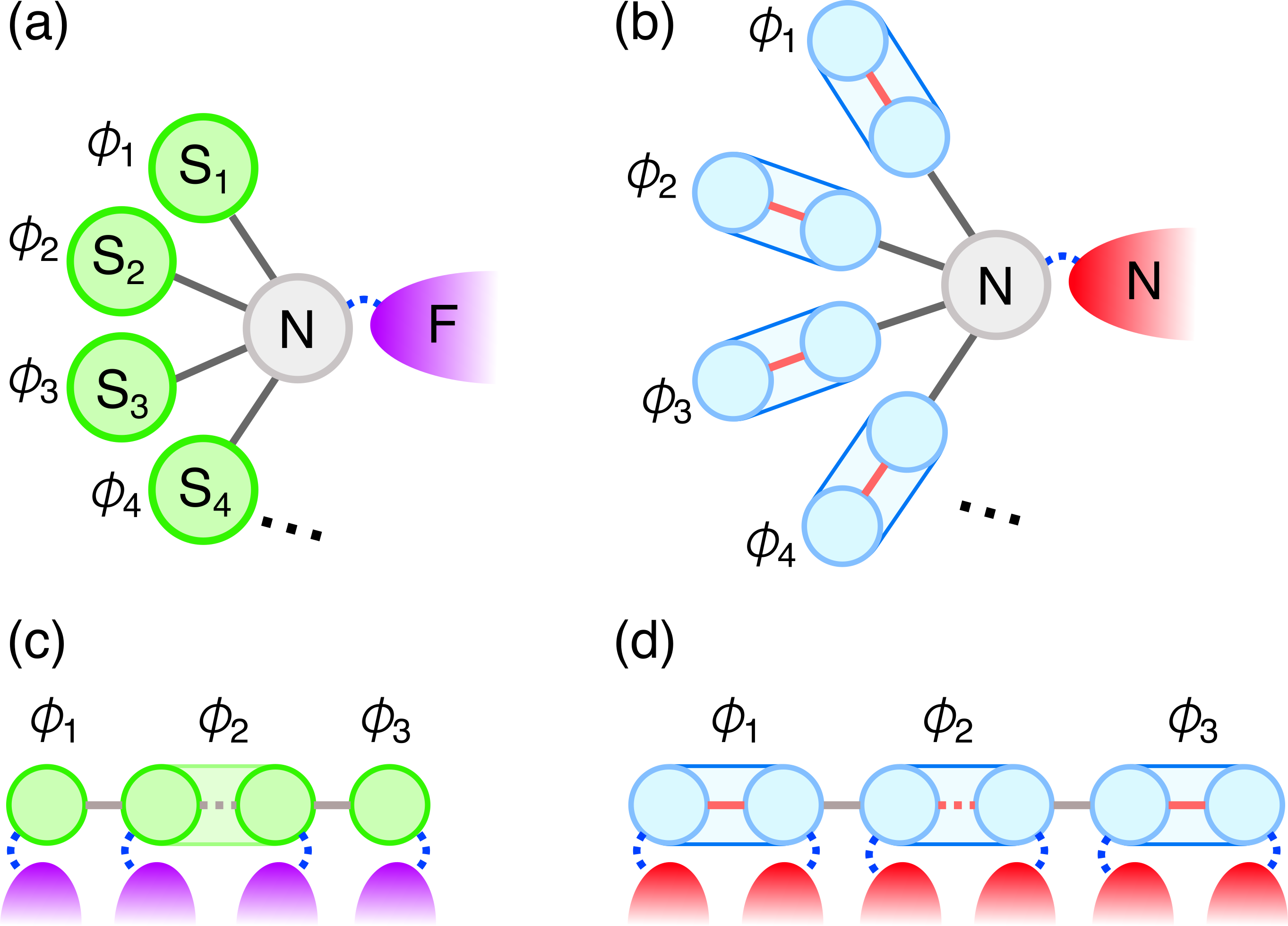}
  	\caption{NH multiterminal JJs. (a) Multiple conventional spin-singlet $s$-wave superconductors (green, S$_{i}$) with distinct phases $\phi_{i}$ are coupled to a normal region (gray, N) which is open to a ferromagnetic lead (magenta, F). (b) The same as in (a) but with two-site  $p$-wave superconductors (cyan boxes with linked circles), known as minimal Kitaev chains,  coupled to a normal region that is open to a normal reservoir (red, N). (c,d) JJs formed by three superconductors of (a,b), open to ferromagnet and normal reservoirs, respectively.}
\label{Fig0} 
\end{figure}

This work investigates NH multiterminal JJs by coupling superconductors with distinct superconducting phases to a normal reservoir; see Fig. \,\ref{Fig0}(a,b). In particular, we focus on two types of multiterminal JJs formed by conventional spin-singlet $s$-wave superconductor and minimal Kitaev chains with spin-triplet $p$-wave superconductivity. We discover that having multiple superconducting phases offers the possibility to realize and control EPs in higher dimensions, originating, for instance, exceptional lines and surfaces connecting stable Fermi arcs in JJs with three and four superconductors. Furthermore, for JJs with laterally coupled superconductors, we find that exceptional points can mediate the formation of hybridized Andreev bound states 
 and produce larger supercurrents that would be tiny otherwise. Our work demonstrates the importance of multiterminal JJs for engineering higher dimensional   NH topological phases.

\section{NH multiterminal JJs}
We first focus on NH JJs where multiple superconductors (Ss) are coupled to the same normal lead via a quantum dot (N); see Fig.\,\ref{Fig0}(a,b). These open systems are modeled by an effective NH Hamiltonian that reads, 
\begin{equation}
\label{Eq1}
H_{\rm eff}(\phi_{i},\omega)=H_{\rm JJ}(\phi_{i})+\Sigma^{r}(\omega)\,,
\end{equation}
where $H_{\rm JJ}(\phi_{i})$ models the JJs with  superconducting phases $\phi_{i}$ and $\Sigma^{r}(\omega)$ is the frequency-dependent retarded self-energy that accounts for coupling to the lead.  Since phase differences matter, when coupling $n$ superconductors, only $n-1$ phases remain in Eq.\,(\ref{Eq1}), which we will exploit when addressing the respective JJs. For pedagogical purposes, we focus on   JJs formed by three and four superconductors, including conventional spin-singlet $s$-wave \cite{Tinkham} and minimal Kitaev chains \cite{PhysRevB.86.134528,cayao2024NHtwositeKitaev} which have spin-triplet $p$-wave superconductivity \cite{cayao2024pairPMMMs,tanaka2024theory}, see below. The choice of  Josephson junctions with distinct number of superconductors is important because it provides an intuitive understanding of the interplay between non-Hermiticity and the multiple superconducting phases for realizing higher dimensional non-Hermitian topological phases. The effective Hamiltonian hosts the particle-hole symmetry, $CH^*_{\rm eff}(\phi_i, \omega)C^{\dagger}=-H_{\rm eff}(\phi_i, -\omega)$, where $C$ is a unitary operator with $CC^*=1$ \cite{cayao2023non}.
Thus, $H^*_{\rm eff}$ can host topologically stable EPs with zero real energy.  Since $\phi_i$ does not change the sign under particle-hole symmetry, the Hamiltonian belongs to class ${\cal P}$D$^\dagger$, as demonstrated in the Supplemental Material of Ref.\,\cite{KBS19}. Thus, from the classification given in Ref.\,\cite{KBS19}, it is very likely that NH multiterminal Josephson junctions will support topologically stable EPs with codimension one, namely EPs in one dimension, exceptional lines in two dimensions, and exceptional surfaces in three dimensions.

A JJ with three superconductors (Ss) is modeled by
\begin{equation}
\label{Eq2}
H_{\rm JJ}=
\begin{pmatrix}
H_{\rm S _{1}}&V^{\dagger}&0&0\\
V&H_{\rm N}&V&V\\
0&V^{\dagger}&H_{\rm S_{2}}&0\\
0&V^{\dagger}&0&H_{\rm S_{3}}
\end{pmatrix}\,,
\end{equation}
where $H_{\rm S_{\alpha}}$ and $V$ model the  Ss and their coupling to N. A conventional  $s$-wave superconductor   S$_{\alpha}$ is modeled by $H^{s}_{\rm S_{\alpha}}=\varepsilon_{\alpha}\tau_{z}+{\rm Re}(\Delta_{\alpha})\sigma_{y}\tau_{y}-{\rm Im}(\Delta_{\alpha})\sigma_{y}\tau_{x}$, with an $s$-wave pair potential $\Delta_{\alpha}=\Delta{\rm e}^{i\phi_{\alpha}}$. Here, $\sigma_i$ and $\tau_i$ are the Pauli matrices in spin and Nambu spaces. Also, $H_{\rm N}=\varepsilon_{\rm N}\tau_{z}$ models the N region, while $V=t\sigma_{0}\tau_{z}$ is the hopping between N and S$_{\alpha}$.  Non-Hermiticity is considered due to coupling N to a ferromagnet lead F [Fig.\,\ref{Fig0}(a)], which induces a spin-dependent self-energy that in the wide-band limit reads \cite{datta1997electronic,PhysRevB.105.094502,cayao2023non} $\Sigma^{r}(\omega=0)={\rm diag}(0,\Sigma_{\rm N}^{r},0,0)$, with $\Sigma_{\rm N}={\rm diag}(\Sigma^{r}_{\rm e},\Sigma^{r}_{\rm h})$ where $\Sigma^{r}_{\rm e,h}=-i \Gamma \sigma_{0}-i\gamma \sigma_{z}$, $\Gamma=(\Gamma_{\uparrow}+\Gamma_{\downarrow})/2$, $\gamma=(\Gamma_{\uparrow}-\Gamma_{\downarrow})/2$;   $\Gamma_{\sigma}$ is the coupling of spin $\sigma$   to F. Similarly, we model a NH JJ with four Ss. It is worth noting that, although ferromagnetism and superconductivity might be seen as antagonistic, moderate values of Zeeman fields  in the ferromagnet leads are expected to give sizeable couplings and not to be detrimental \cite{RevModPhys.77.1321}.

In the case of JJs based on minimal Kitaev chains [Fig.\,\ref{Fig0}(b)], they are modeled by  Eq.\,(\ref{Eq2}) but with two-site $p$-wave Ss given by \cite{PhysRevB.86.134528,cayao2024pairPMMMs} $H^{p}_{\rm S_{\alpha}}=\varepsilon_{\rm 1}\eta_{+}\tau_{z}+\varepsilon_{\rm 2}\eta_{-}\tau_{z}+t_{\alpha}\eta_{x}\tau_{z}+{\rm Re}(\Delta_{\alpha_{p}})\eta_{y}\tau_{y}-{\rm Im}(\Delta_{\alpha_{p}})\eta_{y}\tau_{x}$, where $\varepsilon_{n}$ are the onsite energies, $\Delta_{\rm \alpha_{p}}=\Delta_{\alpha}{\rm e}^{i\phi_{\alpha}}$ is the $p$-wave pair potential  with phase $\phi_{\alpha}$; here $\eta_{\pm}=(\eta_{0}\pm\eta_{z})/2$ and $\eta_{i}$ are   Pauli matrices in the site subspace.    Moreover,  $H_{\rm N}=\varepsilon_{\rm N}\tau_{z}$, while $V=(\tau/2)[\eta_{x}-i\eta_{y},-(\eta_{0}-\eta_{z})]_{2\times4}$. We consider coupling N to a normal lead by $\Sigma_{\rm N}=-i\Gamma_{\rm N}$ to account for NH effects.  Despite the simplicity of our models, they hold particular experimental relevance in the physics of Andreev molecules \cite{Su_2017,Pillet_2019,Coraiola_2023,matsuo2023phase,matsuo2023phase2} and poor man's Majorana modes \cite{dvir2023realization,bordin2023crossed,PhysRevX.13.031031,zatelli2023robustpoorMajo}.  We are interested in EPs forming in NH JJs given by Eq.\,(\ref{Eq2}), especially how the multiple superconducting phases can realize and control them.  Since only phase differences matter, we set $\phi_{1}=0$ and numerically explore EPs in the spaces of $\phi_{2,3}$ and $\phi_{2,3,4}$ for three and four terminal JJs, respectively.

\begin{figure}[!t]
\centering
	\includegraphics[width=0.49\textwidth]{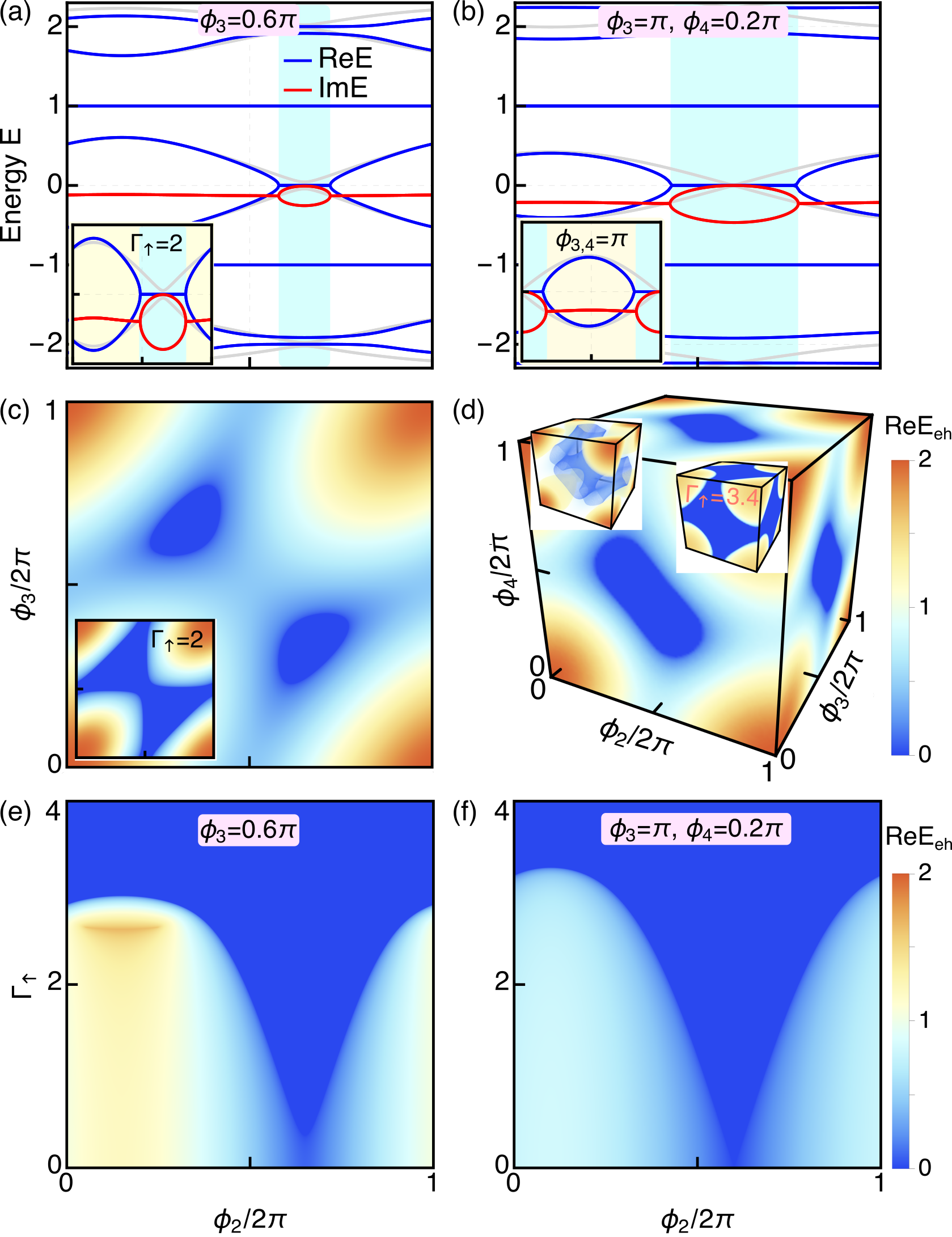}
 	\caption{(a) Re (blue) and Im (red) parts of the eigenvalues of an NH JJ formed by three conventional Ss as a function of $\phi_{2}$ at  $\phi_{3}=0.6\pi$, $\Gamma_{\uparrow}=1$. The inset shows the same as  (a) but for $\Gamma_{\uparrow}=2$.  (b) The same as in (a) but for a JJ with four conventional superconductors at   $\phi_{3}=\pi$, $\phi_{4}=0.2\pi$, $\Gamma_{\uparrow}=2$. The inset shows the same as   (b) but for  $\phi_{3,4}=\pi$. The ends of the cyan regions in (a,b) mark the EPs. (c,d) Re part of the difference between the lowest positive and lowest negative eigenvalues of (a,b) as a function of $\phi_{2,3}$ and $\phi_{2,3,4}$. The insets in (c,d) are for $\Gamma_{\uparrow}=2$ and $\Gamma_{\uparrow}=3.4$, respectively; the left inset in (d) shows the inner side of the blue region. (e,f) Same as in (c,d) as a function of  $\phi_{2}$ and $\Gamma_{\uparrow}$ for the phases of (a,b). The ends of the blue regions mark the EPs.	   Parameters: $\phi_{1}=0$, $\Delta=1$, $t=1$, $\varepsilon_{\alpha}=0$, $\Gamma_{\downarrow}=0$.}
\label{Fig1} 
\end{figure}

\subsection{NH multiterminal JJs based on conventional superconductors}
We start by analyzing NH JJs with three and four conventional Ss; see 
Fig.\,\ref{Fig1}(a,c,e) and Fig.\,\ref{Fig1}(b,d,f), respectively.   Fig.\,\ref{Fig1}(a) shows the Re and Im parts of the energy levels for JJs with three Ss as a function of $\phi_{2}$ at $\phi_{3}=0.6\pi$, $\Gamma_{\uparrow}=1$, while  Fig.\,\ref{Fig1}(b) for JJs with four Ss as a function of $\phi_{2}$ at $\phi_{3}=\pi$, $\phi_{4}=0.2\pi$, $\Gamma_{\uparrow}=1$. In the Hermitian regime ($\Gamma_{\sigma}=0$), the low-energy Andreev spectrum strongly depends on the phases,  with the Andreev bound states (ABSs) developing an asymmetric cosine-like profile for $\phi_{i}=\pi$   that is known to be common in  Hermitian multiterminal JJs \cite{riwar2016multi}, see gray curves in Fig.\,\ref{Fig1}(a,b).  In the NH regime,  the lowest positive and negative energy levels, the ABSs, merge at zero Re energy and remain pinned over a finite range of $\phi_{2}$, see Fig.\,\ref{Fig1}(a,b);  within the zero Re energy line (shaded cyan region), the Im parts split but merge at the ends, where we have verified that the associated wavefunctions become parallel.  Therefore, since eigenvalues and eigenfunctions coalesce at these points, they are EPs \cite{OS23},  notably, here emerging fully controlled by the multiple superconducting phases of the NH JJs with three and four Ss. The regions with zero Re energy and EPs strongly depend on the interplay of non-Hermiticity and the multiple phases; see insets of Fig.\,\ref{Fig1}(a,b). The fact that non-Hermiticity in Fig.\,\ref{Fig1}(a,b) only affects the ABSs, while maintaining a finite energy gap, is a strong indicator that the superconducting gap is not destroyed.

Further insights on how EPs depend on the phases are obtained from the Re part of the difference between lowest positive and lowest negative energies (${\rm Re}E_{\rm eh}$), shown in Fig.\,\ref{Fig1}(c,d) as a function of $\phi_{2,3}$ for the spectra in Fig.\,\ref{Fig1}(a,b). The blue regions indicate ${\rm Re}E_{\rm eh}=0$, which signals the zero Re energy seen in Fig.\,\ref{Fig1}(a,b), and their ends mark the EPs.  At weak non-Hermiticity [Fig.\,\ref{Fig1}(c,d)], the system does not host EPs at $\phi_{2(3)}=0$ and $\phi_{2(3,4)}=0$. Notably, by sweeping all phases $\phi_{\alpha}$ from $0$ to $2\pi$, we find isolated zero Re energy regions (${\rm Re}E_{\rm eh}=0$) with EPs requiring the combined effect of all $\phi_{\alpha}$, see Fig.\,\ref{Fig1}(c,d).  The blue area showing ${\rm Re}E_{\rm eh}=0$ in Fig.\,\ref{Fig1}(c) is a 2D surface with its boundary being a 1D line of EPs (exceptional line), while  ${\rm Re}E_{\rm eh}=0$ in Fig.\,\ref{Fig1}(d)  is a 3D volume with its surface being a 2D region of EPs (exceptional surface).  As non-Hermiticity increases, the isolated blue EP regions merge along $\phi_{2}=-\phi_{3}$ and also extend to $\phi_{2(3)}=0$ and $\phi_{2(3,4)}=0$, see insets of Fig.\,\ref{Fig1}(c,d). The key role of the multiple phases is seen by noting that the blue region with EPs at $\phi_{\alpha}=0$ considerably enlarges as the other phases take finite values.   To avoid confusion, we note that the the blue regions in the faces of the cube in Fig.\,\ref{Fig1}(d) correspond to 2D   regions with ${\rm Re}E_{\rm eh}=0$    which   are connected and hence form a 3D volume with ${\rm Re}E_{\rm eh}=0$.  This connection is depicted in the left inset where we show the skeleton of the right inset which represents the  exceptional surface of the 3D volume. We thus find that the    blue regions in three-terminal JJs  reveal exceptional lines of EPs [Fig.\,\ref{Fig1}(c)], while in four terminal JJs we obtain exceptional surfaces [Fig.\,\ref{Fig1}(d)]. Another interesting feature is that  having multiple phases at strong non-Hermiticity gives larger regions  of ${\rm Re}E_{\rm eh}=0$ along certain axes in the phase space, see axis   $\phi_{2}=-\phi_{3}$ in the inset of Fig.\,\ref{Fig1}(c). The tunability of EPs can be also seen in Fig.\,\ref{Fig1}(e,f), where we present  ${\rm Re}E_{\rm eh}$ as a function of $\Gamma_{\uparrow}$ and $\phi_{2}$ for the spectra of Fig.\,\ref{Fig1}(a,b). Although not presented, we also emphasize that the onsite energies of the Ss can be useful parameters for controlling the formation of EPs \cite{PhysRevB.105.094502,PhysRevB.107.104515}. For realistic conditions, superconductors are large and scalar disorder might be unavoidable; here, scalar disorder modifies the onsite energies, which would then affect the location of the obtained EPs or even allow for EPs due to the interplay of disorder, non-Hermiticity, and the Josephson effect.

     \begin{figure}[!t]
\centering
	\includegraphics[width=0.49\textwidth]{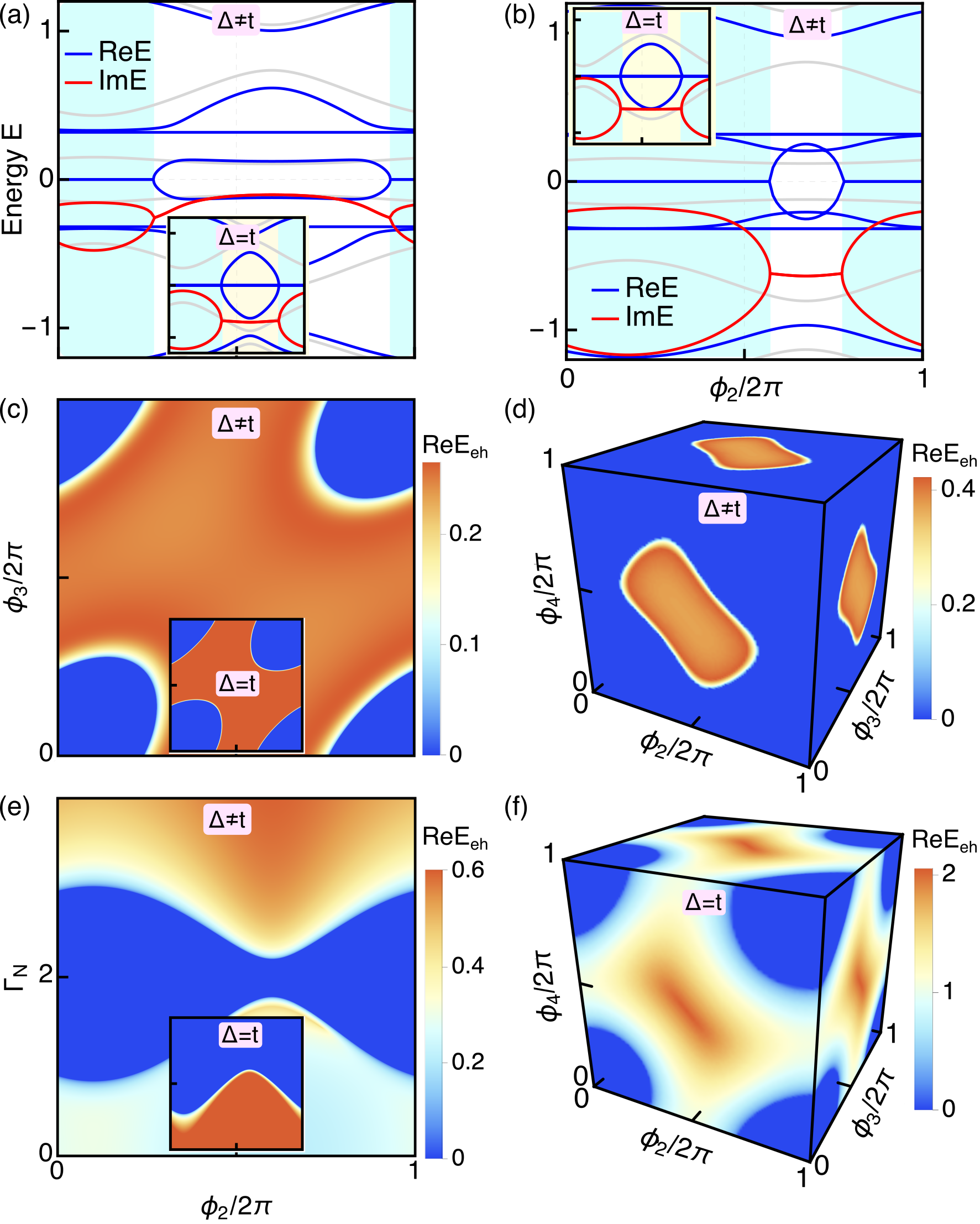}
 	\caption{(a) Re (blue) and Im (red) parts of the eigenvalues of a NH  JJ formed by three minimal Kitaev chains as a function of $\phi_{2}$ at $\Delta=1$, $t=\tau=0.8$, $\phi_{3}=0.4\pi$, $\Gamma_{\rm N}=1$, $\varepsilon_{\alpha}=0.5$.   (b) Same as in (a) but for a JJ with four minimal Kitaev chains at   $\Gamma_{\rm N}=2$, $\varepsilon_{\alpha}=0.5$, $\phi_{3}=0.6\pi$, $\phi_{4}=0.4\pi$. The insets in (a) and (b) show the same as (a) and (b) but for $\phi_{3}=0.4\pi$ and $\phi_{3,4}=0.2\pi$, both cases at $\Delta=t=\tau=1$,	$\Gamma_{\rm N}=2$, $\varepsilon_{\alpha}=0$. The ends of the cyan regions in (a,b) mark the EPs. (c) Re part of the difference between the lowest positive and lowest negative eigenvalues of (a) as a function of $\phi_{2,3}$ at the parameters of (b). The inset shows the same, but for the first excited positive and negative energy levels at the parameters of the inset in (a). (d) The same as (c)  for the eigenvalues of (b) as a function of   $\phi_{2,3,4}$.  (e) The same quantity of (c) as a function of $\phi_{2}$ and $\Gamma_{\rm N}$ for the energies of (a). (f) Same as (d) for the first excited positive and negative energy levels, with the parameters of the inset in (b). The blue areas in (c-f) show levels with zero Re energy whose ends mark the EPs.  Parameters: $\phi_{1}=0$,   $\Gamma_{\downarrow}=0$, $t_{\alpha}=t$.}
\label{Fig2} 
\end{figure}

\subsection{NH multiterminal JJs based on minimal Kitaev chains}
For NH JJs with minimal Kitaev chains,  we distinguish two relevant cases: i) $\Delta=t$, $\varepsilon_{\alpha}=0$, and ii) $\Delta\neq t$, $\varepsilon_{\alpha}\neq0$, both of particular relevance in experiments addressing few-site Kitaev chains \cite{dvir2023realization,bordin2023crossed,PhysRevX.13.031031,zatelli2023robustpoorMajo}. For $\Delta=t$ and $\varepsilon_{\alpha}=0$, the system hosts poor man's Majorana modes in the Hermitian regime \cite{PhysRevB.86.134528}, which are zero energy Majorana modes but without any Hermitian topological protection. For $\Delta\neq t$ and $\varepsilon_{\alpha}\neq0$, the system does not host Majorana-like quasiparticles in the Hermitian regime but it can under the influence of non-Hermiticity, recently shown for a NH minimal Kitaev chain \cite{cayao2024NHtwositeKitaev}. Fig.\,\ref{Fig2}(a,b) shows the Re and Im parts of the energy levels of JJs with three and four Ss as a function of $\phi_{2}$ for finite non-Hermiticity and finite values of $\phi_{\alpha}$ at $\Delta\neq t$, $\varepsilon_{\alpha}\neq0$.  The insets of (a,b) show the Re and Im parts of the energy levels for the respective JJs at $\Delta=t$, $\varepsilon_\alpha=0$; here, the shown Im parts correspond the energy levels that exhibit EPs. The first observation is that non-Hermiticity induces regions with zero Re energy and split Im parts, whose ends mark EPs in JJs with three and four Ss irrespective of $\Delta$ being equal to $t$ or not; see ends of cyan regions in Fig.\,\ref{Fig2}(a,b). For $\Delta\neq t$, EPs form between the lowest positive and lowest negative energies, while  EPs at  $\Delta= t$ form between first positive/negative excited levels.  In both cases, the Re parts develop a circular-like profile between EPs around $\phi_{2}=\pi$, while the Im parts stick to the same value between EPs.  It is worth noting that EPs between first excited levels at $\Delta=t$  emerge because in this case the lowest Re energies  corresponding to the poor mans Majorana modes remain at zero \cite{PhysRevB.86.134528}, depicted by the blue flat line in the insets of Fig.\,\ref{Fig2}(a,b); their Im parts are finite but not shown here. For this reason, the immediate effect of non-Hermiticity at $\Delta=t$  is on the first excited energy levels, which are the ones exhibiting EPs at zero Re energy and protected by NH topology. Since minimal Kitaev chains are small systems, it is unlikely that scalar disorder plays an important role. However, if disorder occurs, additional ingap states might emerge, which can  acquire a phase dependence and then develop EPs.   

The tunability of EPs by the interplay of non-Hermiticity and the distinct phases 
in NH JJs with minimal Kitaev chains can be further seen in Fig.\,\ref{Fig2}(c-f), where we plot the Re part of the difference between levels undergoing EP transitions; the blue regions here indicates having zero Re energy levels with their ends marking EPs. Unlike the EPs in NH JJs with conventional Ss [Fig.\,\ref{Fig1}], here we find that the lowest (first excited) levels closer to $\phi_{\alpha}=0$ are more susceptible to become zero energy and form EPs in an asymmetric fashion with respect to e. g., $\phi_{2}=\pi$ in Fig.\,\ref{Fig2}(a-d,f).  In NH JJs with three Ss [Fig.\,\ref{Fig2}(a,c,e)], the ends of the blue regions mark lines of EPs in the space of $\phi_{2,3}$, while with four Ss such lines become exceptional surfaces as a function of $\phi_{2,3,4}$  [Fig.\,\ref{Fig2}(b,d,f)]. As noted for JJs with four conventional Ss [Fig.\,\ref{Fig1}(d)], the blue regions in the faces of the cube in Fig.\,\ref{Fig2}(d,f)(d,f) are connected and therefore form a 3D volume of zero Re energy; the surface of such volume is a 2D region of EPs or exceptional surface.

We can, therefore, conclude that the interplay of non-Hermiticity and the multiple superconducting phases in multiterminal JJs enables the realization of higher dimensional exceptional regions protected by NH topology with particle-hole symmetry \cite{KBS19}.

\section{NH JJs with three laterally coupled Ss}
Having shown the formation of exceptional lines and surfaces in three and four multiterminal JJs, we now analyze NH JJs with laterally coupled Ss and connected to leads as in   Fig.\,\ref{Fig0}(c,d). These JJs are of experimental relevance in the context of Andreev molecules \cite{Su_2017,matsuo2023phase2,Coraiola_2023} and non-local Josephson effect \cite{Matsuo_2022,matsuo2023phase,Haxell_2023}. We  model these  JJs by
\begin{equation}
\label{Eq3}
H_{\rm JJ}=
\begin{pmatrix}
H_{\rm S_{1}}&V_{12}&0\\
V_{12}^{\dagger}&H_{\rm S_{2}}&V_{23}\\
0&V_{23}^{\dagger}&H_{\rm S_{3}}\\
\end{pmatrix}\,,
\end{equation}
where $H_{\rm S_{\alpha}}$ represents either conventional  Ss or minimal Kitaev chains given below  Eq.\,(\ref{Eq2}),  while $V_{ij}$ is the coupling between  Ss. Since the length of the middle S has shown to be important for assessing Andreev molecules in conventional Ss \cite{Pillet_2019}, for conventional Ss, we consider that $H_{\rm S_{2}}$ is composed of two sites coupled by $V=\tau\sigma_{0}\tau_{z}$: small and large values of $\tau$ then model long and short middle S. Also, $V_{12(23)}=t\sigma_{0}\tau_{z}$.  Non-Hermiticity appears by coupling each S  to leads [Fig.\,\ref{Fig0}(c)], such that a self-energy is added to each $H_{\rm S_{\alpha}}$ in Eq.\,(\ref{Eq3}). The self-energy for conventional Ss is $\Sigma^{r}(\omega=0)={\rm diag}(\Sigma_{1},\Sigma_{2},\Sigma_{3})$, with $\Sigma_{\alpha}={\rm diag}(\Sigma_{\alpha}^{e},\Sigma_{\alpha}^{h})$ and $\Sigma_{\alpha}^{e(h)}=-i \Gamma^{\alpha} \sigma_{0}-i\gamma^{\alpha} \sigma_{z}$, $\Gamma^{\alpha}=(\Gamma^{\alpha}_{\uparrow}+\Gamma^{\alpha}_{\downarrow})/2$, $\gamma^{\alpha}=(\Gamma^{\alpha}_{\uparrow}-\Gamma^{\alpha}_{\downarrow})/2$.  For JJs in Eq.\,(\ref{Eq3}) with minimal Kitaev chains, the couplings are given by $V_{ij}^{\dagger}=t_{ij}(\eta_{x}-i\eta_{y})\tau_{z}$; the self-energy has the same structure as for conventional Ss but in terms of $\eta_{i}$ Pauli matrices instead of $\sigma_{i}$ and   $\Gamma_{n}^{\alpha}$ instead of $\Gamma_{\sigma}^{\alpha}$, with $n=1,2$ denoting the two sites and $\alpha$ labels the left (L), middle (M), and right (R) Ss. By gauge invariance, we take the phase of $H_{\rm S_{2}}$ to be zero, $\phi_{2}=0$, and denote $\phi_{1(3)}=\phi_{\rm L/R}$ as the  phases of the left and right Ss.\\

  \begin{figure}[!t]
\centering
	\includegraphics[width=0.49\textwidth]{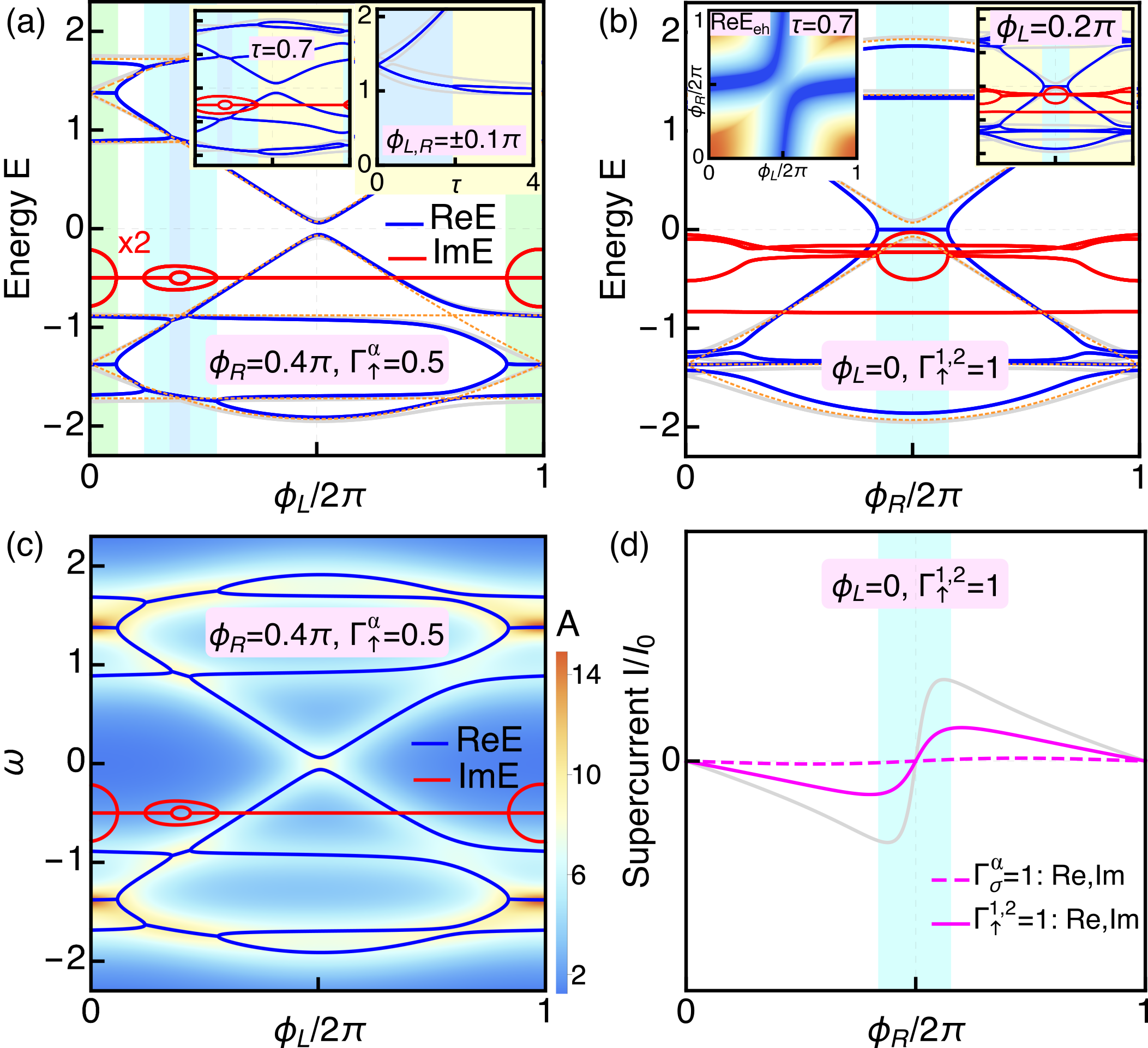}
 	\caption{(a) Re (blue) and Im (red) energy levels as a function of $\phi_{\rm L}$ for a JJ with three laterally coupled conventional Ss at $\Gamma_{\uparrow}^{\alpha}=0.5$, $\phi_{\rm R}=0.4\pi$, $\tau=0.3$, while (b) at $\Gamma_{\uparrow}^{1,2}=1$, $\Gamma_{\uparrow}^{3}=0$,  $\phi_{\rm L}=0$, $\tau=0.3$. Gray curves correspond to the Hermitian regime at the respective $\tau$, while orange dashed curves depict the Hermitian energies with $\tau=0$.  The left inset in (a) shows the same but for $\tau=0.7$, while the right inset shows the positive energies as a function of $\tau$ at $\phi_{\rm L, R}=0.1\pi$. The left inset in (b) is the Re energy difference between the lowest positive and the lowest negative energies as a function of $\phi_{\rm L, R}$ at $\tau=0.7$; the right inset shows the Re and Im energies at $\phi_{\rm L}=0.2\pi$.  In (a,b), the ends of green, cyan, and blue regions mark the EPs.  (c) Spectral function as a function of $\omega$ and $\phi_{\rm L}$ for the regime of (b).   (d) Supercurrents as a function of $\phi_{\rm R}$ at $\phi_{\rm L}=0$ and $\tau=0.3$ for distinct values of non-Hermiticity $\Gamma_{\sigma}^{\alpha}$,   with (solid magenta) and without (dashed magenta) EPs;  without EPs  we set $\Gamma^{\alpha}_{\sigma}=1$.  Parameters: $\phi_{\rm L(R)}=\phi_{\rm 1(3)}$, $\phi_{2}=0$,   $\Gamma^{\alpha}_{\downarrow}=0$, $\Delta=1$, $t=0.93$, $\varepsilon_{\rm S}=0$.
	}
\label{Fig3} 
\end{figure}

\subsection{NH JJs with three laterally coupled conventional superconductors}
We begin by analyzing the NH JJ with three laterally coupled conventional Ss, whose Re and Im energies as a function of  $\phi_{\rm L}$ are plotted in Fig.\,\ref{Fig3}(a) at   $\phi_{\rm R}=0.4\pi$, $\tau=0.3$ and finite non-Hermiticity.  In the Hermitian regime ($\Gamma_{\sigma}^{\alpha}=0$) and with $\tau=0$, the system is composed of two separate JJs:  the   Andreev spectrum dispersing with $\phi_{\rm L}$ corresponds to the left JJ, while the dispersionless levels around $E=\pm1$   come from the right JJ, see orange dashed curves in Fig.\,\ref{Fig3}(a). For $\tau=0.3$, the ABSs hybridize and give rise to an avoided crossing at   $\phi_{L}=\pm\phi_{\rm R}$ that originating the formation of a Hermitian Andreev molecule \cite{Pillet_2019}, see gray curves. A finite non-Hermiticity of the form $\Gamma_{\uparrow}^{\alpha}=0.5$ induces EPs in the Andreev spectrum, which occurs at finite energies between two levels of the same JJ but, notably, also between two levels of distinct JJs. The EPs between levels of the same JJ  appear at the ends of the green shaded region in Fig.\,\ref{Fig3}(a), where the Re parts merge within EPs while the Im parts split. The ends of the cyan and blueish regions mark the EPs between levels of distinct JJs.  The EPs between levels of the same JJ can be removed by increasing $\tau$ but remain robust the EPs between levels of distinct JJs, see left inset of Fig.\,\ref{Fig3}(a).   The effect of $\tau$ can be further elucidated in the right inset of Fig.\,\ref{Fig3}(a), where we plot the positive energies as a function of  $\tau$ at $\phi_{\rm L/R}=\pm0.1\pi$, showing that EPs form between levels of distinct JJs.
An interesting consequence of the EPs found here is that they produce large spectral weights, as obtained in the spectral function in the middle S in Fig.\,\ref{Fig3}(c); the lines connecting EPs produce significant spectral features.  Therefore,  non-Hermiticity can induce and control the formation of  Andreev molecules in JJs with conventional Ss beyond the Hermitian regime.

EPs at zero Re energy are possible by taking another choice of non-Hermiticity. This happens, for instance, when introducing the site-dependent ferromagnetic dissipation $\Gamma_{\uparrow}^{1,2}=1$ and $\Gamma_{\uparrow}^{3}=0$, $\Gamma_{\downarrow}^{\alpha}=0$, with the Re and Im energies plotted in  Fig.\,\ref{Fig3}(b) as a function of $\phi_{\rm R}$ at $\phi_{\rm L}=0$. The ends of the cyan region mark such EPs that occur between the lowest positive and negative energy levels. Notably, these levels correspond to already hybridized ABSs of distinct JJs.  We have verified that these EPs are highly tunable by $\phi_{\rm L/R}$, reflecting the combined effect of non-Hermiticity and the multiple phases; see left and right insets of Fig.\,\ref{Fig3}(b).

Significantly, these EPs at zero Re energy enhance the Josephson current \cite{cayao2023non}: 
Each ABS with quasi-energy $E_n$ carries the supercurrent $I=(-e/\hbar) dRe E_n/d\phi_{\rm L}$,
which diverges at the EPs.
Whereas the Im energy softens the divergence \cite{shen2024nonhermitian, beenakker2024josephson,pino2024thermodynamics}, the current still exhibits the enhancement.
Figure \ref{Fig3}(d) shows the supercurrent $I=(-e\hbar/\pi){\rm Im}\sum_{n}d[E_{n}{\rm ln}(E_{n})]/d\phi_{\rm L}$ \cite{shen2024nonhermitian,beenakker2024josephson, pino2024thermodynamics}, which includes the softening effect. 
For comparison, we also plot the current for the system
with site- and spin-independent dissipation (dashed curve), which does not host  EPs. Although the current with  EPs (solid magenta curve) is less than the corresponding Hermitian case without dissipation (gray curve), it is much larger than the current without EPs because of the EPs.

     \begin{figure}[!t]
\centering
	\includegraphics[width=0.49\textwidth]{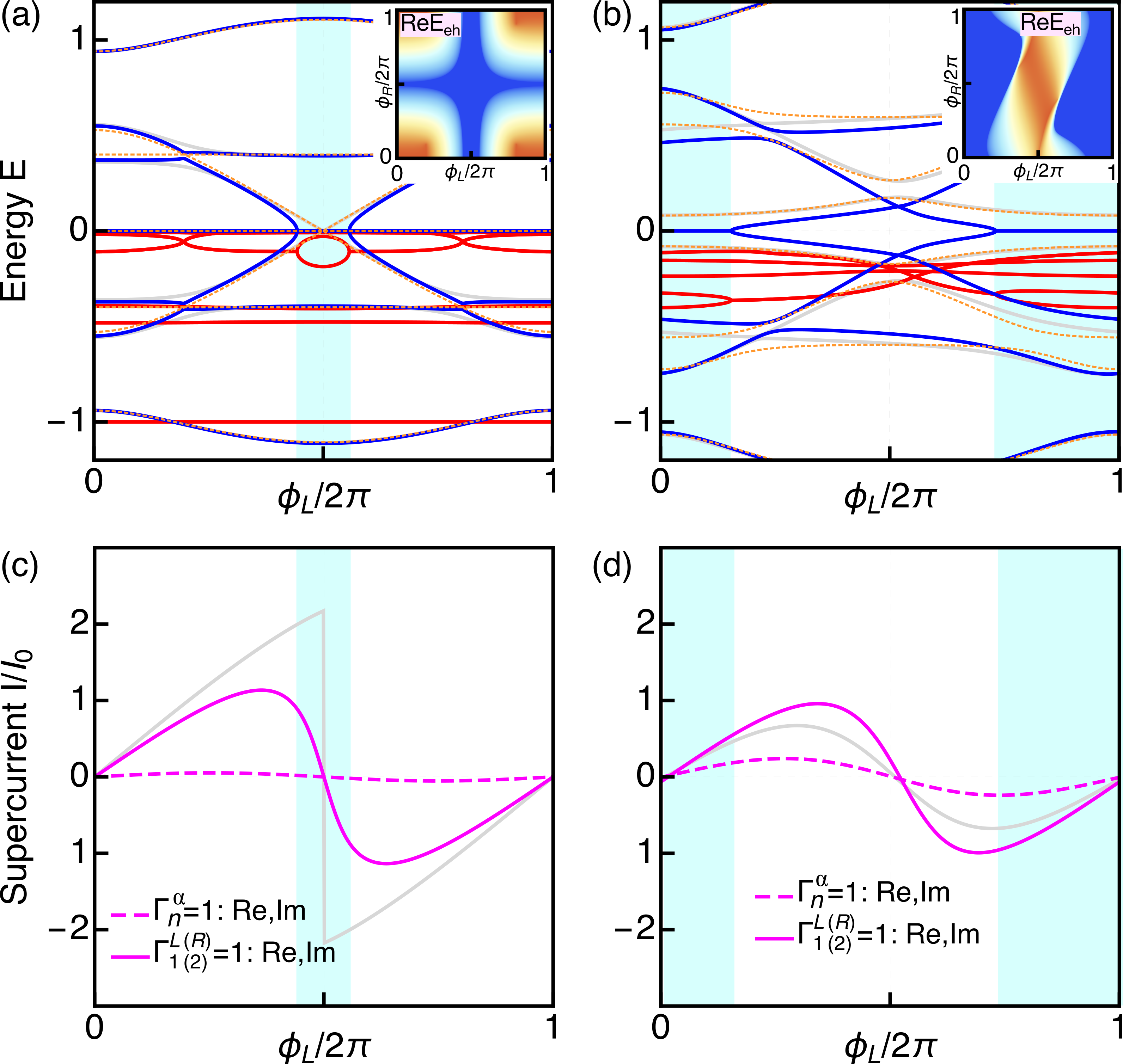}
 	\caption{(a) Re (red) and Im (red) energy levels as a function of $\phi_{\rm L}$ for a JJ with three laterally coupled minimal Kitaev chains at $\Gamma^{\rm L(R)}_{1(2)}=1$, $\phi_{\rm R}=0.2\pi$, $t_{1,3}=\Delta_{1,3}=1$, $t_{2}=\Delta_{2}=0.3$, while (b) at $\Gamma^{\rm L(R)}_{1(2)}=0.5$, $\phi_{\rm R}=0.2\pi$, $t_{1,3}=0.6$, $\Delta_{1,3}=1$, $t_{2}=0.6$, $\Delta_{2}=0.3$. The gray curves correspond to the Hermitian regime at the respective parameters, while orange dashed curves represent the Hermitian energies at $\phi_{\rm R}=0$. The insets show the Re energy difference between the lowest positive and the lowest negative energies as a function of $\phi_{\rm L, R}$.  The ends of the cyan and blue regions mark the EPs. (c,d) Supercurrents as a function of $\phi_{\rm L}$ for (a,b),    with (solid magenta) and without (dashed magenta) EPs;  without EPs we set $\Gamma^{\alpha}_{n}=1$.   Parameters: $\phi_{\rm L(R)}=\phi_{\rm 1(3)}$, $\phi_{2}=0$,   $\Gamma^{\rm L(R)}_{2(1)}=0$,  $\Gamma^{\rm M}_{n}=0$,   $\Delta=1$, $t_{2}=\tau$,  $t_{12}=0.94$, $t_{23}=0.4$, $\varepsilon_{\alpha}=0$.}
\label{Fig4} 
\end{figure}

\subsection{NH JJs with three laterally coupled minimal Kitaev chains}
Regarding JJs with laterally coupled Kitaev chains,  the interplay of non-Hermiticity and $\phi_{\rm L/R}$ provides a controllable way to induce EPs between hybridized ABSs. For instance, in  Fig.\,\ref{Fig4}(a) we show the Re and Im energies as a function of $\phi_{\rm L}$ at  $\phi_{\rm R}$, $t_{2}=0.3$, and $\Gamma^{\rm L(R)}_{1(2)}=1$ when the Hermitian regime hosts poor man's Majorana modes, seen as lines at zero Re energy.  Here,   EPs emerge at the ends of the cyan region due to the coalescence at zero Re energy of the first excited positive and negative ABSs, which contain branches from the left and right JJs and can, therefore, be seen as hybridized ABSs or  Andreev molecules. The hybridzed nature of these NH ABSs can be indeed seen in the orange and gray curves of Fig.\,\ref{Fig4}(a), where we show how the  ABSs at $t_{2}=0$ and $t_{2}=0.3$ appear in the Hermitian regime. EPs can also be induced in the absence of Hermitian poor man's Majorana modes, as we indeed show in Fig.\,\ref{Fig4}(b) for $\Gamma^{\rm L(R)}_{1(2)}=0.5$ and $t_{2}=0.6$. EPs form between the lowest positive and negative hybridized ABSs at zero Re energy and are marked by the ends of the cyan regions in Fig.\,\ref{Fig4}(b). In both cases of Fig.\,\ref{Fig4}(a,b), the emergence of EPs is an effect highly controllable by the interplay of non-Hermiticity and $\phi_{\rm L, R}$, as demonstrated in the respective insets. 

Like the NH JJ with three laterally coupled Ss, the EPs at zero Re energy enhance the Josephson current, as shown in Figs.\,\ref{Fig4} (c,d). Remarkably, the enhanced current can be more significant than the corresponding Hermitian case. See Fig.\,\ref{Fig4} (d). Thus, controlling dissipation produces a measurable, otherwise absent positive impact.

\section{Conclusions}
In conclusion, we study non-Hermitian multiterminal  phase-biased Josephson junctions. We demonstrate the formation of exceptional points entirely controlled by the interplay of non-Hermiticity and multiple superconducting phases. In particular, for Josephson junctions with three and four superconductors, we discover that exceptional points emerge along lines and surfaces as a highly controllable robust effect protected by NH topology. Furthermore, we show that exceptional points can result from hybridized Andreev bound states of distinct Josephson junctions, enhancing spectral signatures and supercurrents.  Our findings are within experimental reach because Josephson junctions similar to the considered here have already been fabricated for studying  Andreev molecules 
\cite{Su_2017,Pillet_2019,Coraiola_2023,matsuo2023phase2}, the  Josephson effect \cite{strambini2016omega,PhysRevX.10.031051,draelos2019supercurrent,graziano2022selective,Matsuo_2022,matsuo2023phase,Haxell_2023},  and poor man's Majorana modes in few-site Kitaev chains \cite{dvir2023realization,bordin2023crossed,PhysRevX.13.031031,zatelli2023robustpoorMajo}. Our work  paves the way for engineering higher dimensional non-Hermitian topological phenomena controlled by the Josephson effect and non-Hermitian topology.

 \emph{Note added.} Recently, a preprint was posted online (Ref.\,\cite{ohnmacht2024NHtopology}), which partially overlaps with some of our results.

\section{Acknowledgements}
 J. C. acknowledges financial support from the Swedish Research Council (Vetenskapsr{\aa}det Grant No. 2021-04121),  the Royal Swedish Academy of Sciences (Grant No. PH2022-0003), and the Carl Trygger's Foundation (Grant No. 22: 2093). M. S. was supported by JST CREST Grant No.\,JPMJCR19T2 and JSPS KAKENHI Grant No.\,JP24K00569. The computations were enabled by resources provided by the National Academic Infrastructure for Supercomputing in Sweden (NAISS), partially funded by the Swedish Research Council through grant agreement no. 2022-06725

\bibliography{biblio}

\begin{thebibliography}{60}%
\makeatletter
\providecommand \@ifxundefined [1]{%
 \@ifx{#1\undefined}
}%
\providecommand \@ifnum [1]{%
 \ifnum #1\expandafter \@firstoftwo
 \else \expandafter \@secondoftwo
 \fi
}%
\providecommand \@ifx [1]{%
 \ifx #1\expandafter \@firstoftwo
 \else \expandafter \@secondoftwo
 \fi
}%
\providecommand \natexlab [1]{#1}%
\providecommand \enquote  [1]{``#1''}%
\providecommand \bibnamefont  [1]{#1}%
\providecommand \bibfnamefont [1]{#1}%
\providecommand \citenamefont [1]{#1}%
\providecommand \href@noop [0]{\@secondoftwo}%
\providecommand \href [0]{\begingroup \@sanitize@url \@href}%
\providecommand \@href[1]{\@@startlink{#1}\@@href}%
\providecommand \@@href[1]{\endgroup#1\@@endlink}%
\providecommand \@sanitize@url [0]{\catcode `\\12\catcode `\$12\catcode
  `\&12\catcode `\#12\catcode `\^12\catcode `\_12\catcode `\%12\relax}%
\providecommand \@@startlink[1]{}%
\providecommand \@@endlink[0]{}%
\providecommand \url  [0]{\begingroup\@sanitize@url \@url }%
\providecommand \@url [1]{\endgroup\@href {#1}{\urlprefix }}%
\providecommand \urlprefix  [0]{URL }%
\providecommand \Eprint [0]{\href }%
\providecommand \doibase [0]{https://doi.org/}%
\providecommand \selectlanguage [0]{\@gobble}%
\providecommand \bibinfo  [0]{\@secondoftwo}%
\providecommand \bibfield  [0]{\@secondoftwo}%
\providecommand \translation [1]{[#1]}%
\providecommand \BibitemOpen [0]{}%
\providecommand \bibitemStop [0]{}%
\providecommand \bibitemNoStop [0]{.\EOS\space}%
\providecommand \EOS [0]{\spacefactor3000\relax}%
\providecommand \BibitemShut  [1]{\csname bibitem#1\endcsname}%
\let\auto@bib@innerbib\@empty
\bibitem [{\citenamefont {El-Ganainy}\ \emph {et~al.}(2018)\citenamefont
  {El-Ganainy}, \citenamefont {Makris}, \citenamefont {Khajavikhan},
  \citenamefont {Musslimani}, \citenamefont {Rotter},\ and\ \citenamefont
  {Christodoulides}}]{el2018non}%
  \BibitemOpen
  \bibfield  {author} {\bibinfo {author} {\bibfnamefont {R.}~\bibnamefont
  {El-Ganainy}}, \bibinfo {author} {\bibfnamefont {K.~G.}\ \bibnamefont
  {Makris}}, \bibinfo {author} {\bibfnamefont {M.}~\bibnamefont {Khajavikhan}},
  \bibinfo {author} {\bibfnamefont {Z.~H.}\ \bibnamefont {Musslimani}},
  \bibinfo {author} {\bibfnamefont {S.}~\bibnamefont {Rotter}},\ and\ \bibinfo
  {author} {\bibfnamefont {D.~N.}\ \bibnamefont {Christodoulides}},\ }\bibfield
   {title} {\bibinfo {title} {Non-{H}ermitian physics and {PT} symmetry},\
  }\href {https://www.nature.com/articles/nphys4323} {\bibfield  {journal}
  {\bibinfo  {journal} {Nat. Phys.}\ }\textbf {\bibinfo {volume} {14}},\
  \bibinfo {pages} {11} (\bibinfo {year} {2018})}\BibitemShut {NoStop}%
\bibitem [{\citenamefont {{\"O}zdemir}\ \emph {et~al.}(2019)\citenamefont
  {{\"O}zdemir}, \citenamefont {Rotter}, \citenamefont {Nori},\ and\
  \citenamefont {Yang}}]{ozdemir2019parity}%
  \BibitemOpen
  \bibfield  {author} {\bibinfo {author} {\bibfnamefont {{\c{S}}.~K.}\
  \bibnamefont {{\"O}zdemir}}, \bibinfo {author} {\bibfnamefont
  {S.}~\bibnamefont {Rotter}}, \bibinfo {author} {\bibfnamefont
  {F.}~\bibnamefont {Nori}},\ and\ \bibinfo {author} {\bibfnamefont
  {L.}~\bibnamefont {Yang}},\ }\bibfield  {title} {\bibinfo {title}
  {Parity--time symmetry and exceptional points in photonics},\ }\href
  {https://www.nature.com/articles/s41563-019-0304-9} {\bibfield  {journal}
  {\bibinfo  {journal} {Nat. Mater.}\ }\textbf {\bibinfo {volume} {18}},\
  \bibinfo {pages} {783} (\bibinfo {year} {2019})}\BibitemShut {NoStop}%
\bibitem [{\citenamefont {Ashida}\ \emph {et~al.}(2020)\citenamefont {Ashida},
  \citenamefont {Gong},\ and\ \citenamefont
  {Ueda}}]{doi:10.1080/00018732.2021.1876991}%
  \BibitemOpen
  \bibfield  {author} {\bibinfo {author} {\bibfnamefont {Y.}~\bibnamefont
  {Ashida}}, \bibinfo {author} {\bibfnamefont {Z.}~\bibnamefont {Gong}},\ and\
  \bibinfo {author} {\bibfnamefont {M.}~\bibnamefont {Ueda}},\ }\bibfield
  {title} {\bibinfo {title} {Non-{H}ermitian physics},\ }\href
  {https://doi.org/10.1080/00018732.2021.1876991} {\bibfield  {journal}
  {\bibinfo  {journal} {Adv. Phys.}\ }\textbf {\bibinfo {volume} {69}},\
  \bibinfo {pages} {249} (\bibinfo {year} {2020})}\BibitemShut {NoStop}%
\bibitem [{\citenamefont {Gong}\ \emph {et~al.}(2018)\citenamefont {Gong},
  \citenamefont {Ashida}, \citenamefont {Kawabata}, \citenamefont {Takasan},
  \citenamefont {Higashikawa},\ and\ \citenamefont {Ueda}}]{PhysRevX.8.031079}%
  \BibitemOpen
  \bibfield  {author} {\bibinfo {author} {\bibfnamefont {Z.}~\bibnamefont
  {Gong}}, \bibinfo {author} {\bibfnamefont {Y.}~\bibnamefont {Ashida}},
  \bibinfo {author} {\bibfnamefont {K.}~\bibnamefont {Kawabata}}, \bibinfo
  {author} {\bibfnamefont {K.}~\bibnamefont {Takasan}}, \bibinfo {author}
  {\bibfnamefont {S.}~\bibnamefont {Higashikawa}},\ and\ \bibinfo {author}
  {\bibfnamefont {M.}~\bibnamefont {Ueda}},\ }\bibfield  {title} {\bibinfo
  {title} {Topological phases of non-{H}ermitian systems},\ }\href
  {https://doi.org/10.1103/PhysRevX.8.031079} {\bibfield  {journal} {\bibinfo
  {journal} {Phys. Rev. X}\ }\textbf {\bibinfo {volume} {8}},\ \bibinfo {pages}
  {031079} (\bibinfo {year} {2018})}\BibitemShut {NoStop}%
\bibitem [{\citenamefont {Kawabata}\ \emph
  {et~al.}(2019{\natexlab{a}})\citenamefont {Kawabata}, \citenamefont
  {Shiozaki}, \citenamefont {Ueda},\ and\ \citenamefont
  {Sato}}]{PhysRevX.9.041015}%
  \BibitemOpen
  \bibfield  {author} {\bibinfo {author} {\bibfnamefont {K.}~\bibnamefont
  {Kawabata}}, \bibinfo {author} {\bibfnamefont {K.}~\bibnamefont {Shiozaki}},
  \bibinfo {author} {\bibfnamefont {M.}~\bibnamefont {Ueda}},\ and\ \bibinfo
  {author} {\bibfnamefont {M.}~\bibnamefont {Sato}},\ }\bibfield  {title}
  {\bibinfo {title} {Symmetry and topology in non-hermitian physics},\ }\href
  {https://doi.org/10.1103/PhysRevX.9.041015} {\bibfield  {journal} {\bibinfo
  {journal} {Phys. Rev. X}\ }\textbf {\bibinfo {volume} {9}},\ \bibinfo {pages}
  {041015} (\bibinfo {year} {2019}{\natexlab{a}})}\BibitemShut {NoStop}%
\bibitem [{\citenamefont {Okuma}\ and\ \citenamefont {Sato}(2023)}]{OS23}%
  \BibitemOpen
  \bibfield  {author} {\bibinfo {author} {\bibfnamefont {N.}~\bibnamefont
  {Okuma}}\ and\ \bibinfo {author} {\bibfnamefont {M.}~\bibnamefont {Sato}},\
  }\bibfield  {title} {\bibinfo {title} {Non-{H}ermitian topological phenomena:
  a review},\ }\href@noop {} {\bibfield  {journal} {\bibinfo  {journal} {Annu.
  Rev. Condens. Matter Phys.}\ ,\ \bibinfo {pages} {83}} (\bibinfo {year}
  {2023})}\BibitemShut {NoStop}%
\bibitem [{\citenamefont {Wiersig}(2020)}]{wiersig2020review}%
  \BibitemOpen
  \bibfield  {author} {\bibinfo {author} {\bibfnamefont {J.}~\bibnamefont
  {Wiersig}},\ }\bibfield  {title} {\bibinfo {title} {Review of exceptional
  point-based sensors},\ }\href
  {https://opg.optica.org/abstract.cfm?uri=prj-8-9-1457} {\bibfield  {journal}
  {\bibinfo  {journal} {Photonics Res.}\ }\textbf {\bibinfo {volume} {8}},\
  \bibinfo {pages} {1457} (\bibinfo {year} {2020})}\BibitemShut {NoStop}%
\bibitem [{\citenamefont {Parto}\ \emph {et~al.}(2020)\citenamefont {Parto},
  \citenamefont {Liu}, \citenamefont {Bahari}, \citenamefont {Khajavikhan},\
  and\ \citenamefont {Christodoulides}}]{parto2020non}%
  \BibitemOpen
  \bibfield  {author} {\bibinfo {author} {\bibfnamefont {M.}~\bibnamefont
  {Parto}}, \bibinfo {author} {\bibfnamefont {Y.~G.}\ \bibnamefont {Liu}},
  \bibinfo {author} {\bibfnamefont {B.}~\bibnamefont {Bahari}}, \bibinfo
  {author} {\bibfnamefont {M.}~\bibnamefont {Khajavikhan}},\ and\ \bibinfo
  {author} {\bibfnamefont {D.~N.}\ \bibnamefont {Christodoulides}},\ }\bibfield
   {title} {\bibinfo {title} {Non-{H}ermitian and topological photonics: optics
  at an exceptional point},\ }\href {https://doi.org/10.1515/nanoph-2020-0434}
  {\bibfield  {journal} {\bibinfo  {journal} {Nanophotonics}\ }\textbf
  {\bibinfo {volume} {10}},\ \bibinfo {pages} {403} (\bibinfo {year}
  {2020})}\BibitemShut {NoStop}%
\bibitem [{\citenamefont {Bergholtz}\ \emph {et~al.}(2021)\citenamefont
  {Bergholtz}, \citenamefont {Budich},\ and\ \citenamefont
  {Kunst}}]{RevModPhys.93.015005}%
  \BibitemOpen
  \bibfield  {author} {\bibinfo {author} {\bibfnamefont {E.~J.}\ \bibnamefont
  {Bergholtz}}, \bibinfo {author} {\bibfnamefont {J.~C.}\ \bibnamefont
  {Budich}},\ and\ \bibinfo {author} {\bibfnamefont {F.~K.}\ \bibnamefont
  {Kunst}},\ }\bibfield  {title} {\bibinfo {title} {Exceptional topology of
  non-{H}ermitian systems},\ }\href
  {https://doi.org/10.1103/RevModPhys.93.015005} {\bibfield  {journal}
  {\bibinfo  {journal} {Rev. Mod. Phys.}\ }\textbf {\bibinfo {volume} {93}},\
  \bibinfo {pages} {015005} (\bibinfo {year} {2021})}\BibitemShut {NoStop}%
\bibitem [{\citenamefont {Bessho}\ \emph {et~al.}(2019)\citenamefont {Bessho},
  \citenamefont {Kawabata},\ and\ \citenamefont
  {Sato}}]{doi:10.7566/JPSCP.30.011098}%
  \BibitemOpen
  \bibfield  {author} {\bibinfo {author} {\bibfnamefont {T.}~\bibnamefont
  {Bessho}}, \bibinfo {author} {\bibfnamefont {K.}~\bibnamefont {Kawabata}},\
  and\ \bibinfo {author} {\bibfnamefont {M.}~\bibnamefont {Sato}},\ }\bibinfo
  {title} {Topological classificaton of non-{H}ermitian gapless phases:
  Exceptional points and bulk fermi arcs},\ in\ \href
  {https://doi.org/10.7566/JPSCP.30.011098} {\emph {\bibinfo {booktitle} {Proc.
  Int. Conf. on Strongly Correlated Electron Systems (SCES2019)}}}\ (\bibinfo
  {publisher} {Physical Society of Japan},\ \bibinfo {year} {2019})\
  Chap.~\bibinfo {chapter} {30}, p.\ \bibinfo {pages} {011098}\BibitemShut
  {NoStop}%
\bibitem [{\citenamefont {Meden}\ \emph {et~al.}(2023)\citenamefont {Meden},
  \citenamefont {Grunwald},\ and\ \citenamefont {Kennes}}]{Meden_2023}%
  \BibitemOpen
  \bibfield  {author} {\bibinfo {author} {\bibfnamefont {V.}~\bibnamefont
  {Meden}}, \bibinfo {author} {\bibfnamefont {L.}~\bibnamefont {Grunwald}},\
  and\ \bibinfo {author} {\bibfnamefont {D.~M.}\ \bibnamefont {Kennes}},\
  }\bibfield  {title} {\bibinfo {title} {$\mathcal{PT}$-symmetric,
  non-{H}ermitian quantum many-body physics--a methodological perspective},\
  }\href {https://dx.doi.org/10.1088/1361-6633/ad05f3} {\bibfield  {journal}
  {\bibinfo  {journal} {Rep. Prog. Phys.}\ }\textbf {\bibinfo {volume} {86}},\
  \bibinfo {pages} {124501} (\bibinfo {year} {2023})}\BibitemShut {NoStop}%
\bibitem [{\citenamefont {Datta}(1997)}]{datta1997electronic}%
  \BibitemOpen
  \bibfield  {author} {\bibinfo {author} {\bibfnamefont {S.}~\bibnamefont
  {Datta}},\ }\href@noop {} {\emph {\bibinfo {title} {Electronic transport in
  mesoscopic systems}}}\ (\bibinfo  {publisher} {Cambridge university press},\
  \bibinfo {year} {1997})\BibitemShut {NoStop}%
\bibitem [{\citenamefont {Pikulin}\ and\ \citenamefont
  {Nazarov}(2012)}]{pikulin2012topological}%
  \BibitemOpen
  \bibfield  {author} {\bibinfo {author} {\bibfnamefont {D.~I.}\ \bibnamefont
  {Pikulin}}\ and\ \bibinfo {author} {\bibfnamefont {Y.~V.}\ \bibnamefont
  {Nazarov}},\ }\bibfield  {title} {\bibinfo {title} {Topological properties of
  superconducting junctions},\ }\href@noop {} {\bibfield  {journal} {\bibinfo
  {journal} {JETP Lett.}\ }\textbf {\bibinfo {volume} {94}},\ \bibinfo {pages}
  {693} (\bibinfo {year} {2012})}\BibitemShut {NoStop}%
\bibitem [{\citenamefont {Pikulin}\ and\ \citenamefont
  {Nazarov}(2013)}]{PhysRevB.87.235421}%
  \BibitemOpen
  \bibfield  {author} {\bibinfo {author} {\bibfnamefont {D.~I.}\ \bibnamefont
  {Pikulin}}\ and\ \bibinfo {author} {\bibfnamefont {Y.~V.}\ \bibnamefont
  {Nazarov}},\ }\bibfield  {title} {\bibinfo {title} {Two types of topological
  transitions in finite {M}ajorana wires},\ }\href
  {https://doi.org/10.1103/PhysRevB.87.235421} {\bibfield  {journal} {\bibinfo
  {journal} {Phys. Rev. B}\ }\textbf {\bibinfo {volume} {87}},\ \bibinfo
  {pages} {235421} (\bibinfo {year} {2013})}\BibitemShut {NoStop}%
\bibitem [{\citenamefont {Ioselevich}\ and\ \citenamefont
  {Feigel'man}(2013)}]{Ioselevich_2013}%
  \BibitemOpen
  \bibfield  {author} {\bibinfo {author} {\bibfnamefont {P.~A.}\ \bibnamefont
  {Ioselevich}}\ and\ \bibinfo {author} {\bibfnamefont {M.~V.}\ \bibnamefont
  {Feigel'man}},\ }\bibfield  {title} {\bibinfo {title} {Tunneling conductance
  due to a discrete spectrum of {A}ndreev states},\ }\href
  {https://dx.doi.org/10.1088/1367-2630/15/5/055011} {\bibfield  {journal}
  {\bibinfo  {journal} {New J. Phys.}\ }\textbf {\bibinfo {volume} {15}},\
  \bibinfo {pages} {055011} (\bibinfo {year} {2013})}\BibitemShut {NoStop}%
\bibitem [{\citenamefont {San-Jos\'{e}}\ \emph {et~al.}(2016)\citenamefont
  {San-Jos\'{e}}, \citenamefont {Cayao}, \citenamefont {Prada},\ and\
  \citenamefont {Aguado}}]{JorgeEPs}%
  \BibitemOpen
  \bibfield  {author} {\bibinfo {author} {\bibfnamefont {P.}~\bibnamefont
  {San-Jos\'{e}}}, \bibinfo {author} {\bibfnamefont {J.}~\bibnamefont {Cayao}},
  \bibinfo {author} {\bibfnamefont {E.}~\bibnamefont {Prada}},\ and\ \bibinfo
  {author} {\bibfnamefont {R.}~\bibnamefont {Aguado}},\ }\bibfield  {title}
  {\bibinfo {title} {Majorana bound states from exceptional points in
  non-topological superconductors},\ }\href
  {http://dx.doi.org/10.1038/srep21427} {\bibfield  {journal} {\bibinfo
  {journal} {Sci. Rep.}\ }\textbf {\bibinfo {volume} {6}},\ \bibinfo {pages}
  {21427} (\bibinfo {year} {2016})}\BibitemShut {NoStop}%
\bibitem [{\citenamefont {Avila}\ \emph {et~al.}(2019)\citenamefont {Avila},
  \citenamefont {Pe{\~n}aranda}, \citenamefont {Prada}, \citenamefont
  {San-Jose},\ and\ \citenamefont {Aguado}}]{avila2019non}%
  \BibitemOpen
  \bibfield  {author} {\bibinfo {author} {\bibfnamefont {J.}~\bibnamefont
  {Avila}}, \bibinfo {author} {\bibfnamefont {F.}~\bibnamefont
  {Pe{\~n}aranda}}, \bibinfo {author} {\bibfnamefont {E.}~\bibnamefont
  {Prada}}, \bibinfo {author} {\bibfnamefont {P.}~\bibnamefont {San-Jose}},\
  and\ \bibinfo {author} {\bibfnamefont {R.}~\bibnamefont {Aguado}},\
  }\bibfield  {title} {\bibinfo {title} {Non-{H}ermitian topology as a unifying
  framework for the {A}ndreev versus {M}ajorana states controversy},\
  }\href@noop {} {\bibfield  {journal} {\bibinfo  {journal} {Commun. Phys.}\
  }\textbf {\bibinfo {volume} {2}},\ \bibinfo {pages} {133} (\bibinfo {year}
  {2019})}\BibitemShut {NoStop}%
\bibitem [{\citenamefont {Cayao}\ and\ \citenamefont
  {Black-Schaffer}(2022)}]{PhysRevB.105.094502}%
  \BibitemOpen
  \bibfield  {author} {\bibinfo {author} {\bibfnamefont {J.}~\bibnamefont
  {Cayao}}\ and\ \bibinfo {author} {\bibfnamefont {A.~M.}\ \bibnamefont
  {Black-Schaffer}},\ }\bibfield  {title} {\bibinfo {title} {Exceptional
  odd-frequency pairing in non-{H}ermitian superconducting systems},\ }\href
  {https://doi.org/10.1103/PhysRevB.105.094502} {\bibfield  {journal} {\bibinfo
   {journal} {Phys. Rev. B}\ }\textbf {\bibinfo {volume} {105}},\ \bibinfo
  {pages} {094502} (\bibinfo {year} {2022})}\BibitemShut {NoStop}%
\bibitem [{\citenamefont {Javed}\ \emph {et~al.}(2023)\citenamefont {Javed},
  \citenamefont {Schwibbert},\ and\ \citenamefont
  {Riwar}}]{PhysRevB.107.035408}%
  \BibitemOpen
  \bibfield  {author} {\bibinfo {author} {\bibfnamefont {M.~A.}\ \bibnamefont
  {Javed}}, \bibinfo {author} {\bibfnamefont {J.}~\bibnamefont {Schwibbert}},\
  and\ \bibinfo {author} {\bibfnamefont {R.-P.}\ \bibnamefont {Riwar}},\
  }\bibfield  {title} {\bibinfo {title} {Fractional {J}osephson effect versus
  fractional charge in superconducting--normal metal hybrid circuits},\ }\href
  {https://doi.org/10.1103/PhysRevB.107.035408} {\bibfield  {journal} {\bibinfo
   {journal} {Phys. Rev. B}\ }\textbf {\bibinfo {volume} {107}},\ \bibinfo
  {pages} {035408} (\bibinfo {year} {2023})}\BibitemShut {NoStop}%
\bibitem [{\citenamefont {Cayao}\ and\ \citenamefont
  {Black-Schaffer}(2023)}]{PhysRevB.107.104515}%
  \BibitemOpen
  \bibfield  {author} {\bibinfo {author} {\bibfnamefont {J.}~\bibnamefont
  {Cayao}}\ and\ \bibinfo {author} {\bibfnamefont {A.~M.}\ \bibnamefont
  {Black-Schaffer}},\ }\bibfield  {title} {\bibinfo {title} {Bulk {B}ogoliubov
  {F}ermi arcs in non-{H}ermitian superconducting systems},\ }\href
  {https://doi.org/10.1103/PhysRevB.107.104515} {\bibfield  {journal} {\bibinfo
   {journal} {Phys. Rev. B}\ }\textbf {\bibinfo {volume} {107}},\ \bibinfo
  {pages} {104515} (\bibinfo {year} {2023})}\BibitemShut {NoStop}%
\bibitem [{\citenamefont {Kornich}\ and\ \citenamefont
  {Trauzettel}(2022)}]{PhysRevResearch.4.L022018}%
  \BibitemOpen
  \bibfield  {author} {\bibinfo {author} {\bibfnamefont {V.}~\bibnamefont
  {Kornich}}\ and\ \bibinfo {author} {\bibfnamefont {B.}~\bibnamefont
  {Trauzettel}},\ }\bibfield  {title} {\bibinfo {title} {Signature of
  $\mathcal{P}\mathcal{T}$-symmetric non-{H}ermitian superconductivity in
  angle-resolved photoelectron fluctuation spectroscopy},\ }\href
  {https://doi.org/10.1103/PhysRevResearch.4.L022018} {\bibfield  {journal}
  {\bibinfo  {journal} {Phys. Rev. Research}\ }\textbf {\bibinfo {volume}
  {4}},\ \bibinfo {pages} {L022018} (\bibinfo {year} {2022})}\BibitemShut
  {NoStop}%
\bibitem [{\citenamefont {M\'elin}(2022)}]{PhysRevB.105.155418}%
  \BibitemOpen
  \bibfield  {author} {\bibinfo {author} {\bibfnamefont {R.}~\bibnamefont
  {M\'elin}},\ }\bibfield  {title} {\bibinfo {title} {Multiterminal ballistic
  {J}osephson junctions coupled to normal leads},\ }\href
  {https://doi.org/10.1103/PhysRevB.105.155418} {\bibfield  {journal} {\bibinfo
   {journal} {Phys. Rev. B}\ }\textbf {\bibinfo {volume} {105}},\ \bibinfo
  {pages} {155418} (\bibinfo {year} {2022})}\BibitemShut {NoStop}%
\bibitem [{\citenamefont {Cayao}(2024{\natexlab{a}})}]{cayao2024nonhermitian}%
  \BibitemOpen
  \bibfield  {author} {\bibinfo {author} {\bibfnamefont {J.}~\bibnamefont
  {Cayao}},\ }\bibfield  {title} {\bibinfo {title} {Non-{H}ermitian zero-energy
  pinning of {A}ndreev and {M}ajorana bound states in
  superconductor-semiconductor systems},\ }\href
  {https://doi.org/10.1103/PhysRevB.110.085414} {\bibfield  {journal} {\bibinfo
   {journal} {Phys. Rev. B}\ }\textbf {\bibinfo {volume} {110}},\ \bibinfo
  {pages} {085414} (\bibinfo {year} {2024}{\natexlab{a}})}\BibitemShut
  {NoStop}%
\bibitem [{\citenamefont {Ezawa}(2024)}]{PhysRevB.109.L161404}%
  \BibitemOpen
  \bibfield  {author} {\bibinfo {author} {\bibfnamefont {M.}~\bibnamefont
  {Ezawa}},\ }\bibfield  {title} {\bibinfo {title} {Even-odd effect on
  robustness of {M}ajorana edge states in short {K}itaev chains},\ }\href
  {https://doi.org/10.1103/PhysRevB.109.L161404} {\bibfield  {journal}
  {\bibinfo  {journal} {Phys. Rev. B}\ }\textbf {\bibinfo {volume} {109}},\
  \bibinfo {pages} {L161404} (\bibinfo {year} {2024})}\BibitemShut {NoStop}%
\bibitem [{\citenamefont {Cayao}\ and\ \citenamefont
  {Sato}(2024)}]{cayao2023non}%
  \BibitemOpen
  \bibfield  {author} {\bibinfo {author} {\bibfnamefont {J.}~\bibnamefont
  {Cayao}}\ and\ \bibinfo {author} {\bibfnamefont {M.}~\bibnamefont {Sato}},\
  }\bibfield  {title} {\bibinfo {title} {Non-{H}ermitian phase-biased
  {J}osephson junctions},\ }\href
  {https://doi.org/10.1103/PhysRevB.110.L201403} {\bibfield  {journal}
  {\bibinfo  {journal} {Phys. Rev. B}\ }\textbf {\bibinfo {volume} {110}},\
  \bibinfo {pages} {L201403} (\bibinfo {year} {2024})}\BibitemShut {NoStop}%
\bibitem [{\citenamefont {Li}\ \emph {et~al.}(2023)\citenamefont {Li},
  \citenamefont {Sun},\ and\ \citenamefont {Trauzettel}}]{li2023anomalous}%
  \BibitemOpen
  \bibfield  {author} {\bibinfo {author} {\bibfnamefont {C.-A.}\ \bibnamefont
  {Li}}, \bibinfo {author} {\bibfnamefont {H.-P.}\ \bibnamefont {Sun}},\ and\
  \bibinfo {author} {\bibfnamefont {B.}~\bibnamefont {Trauzettel}},\ }\bibfield
   {title} {\bibinfo {title} {Anomalous {A}ndreev bound states in
  non-{H}ermitian {J}osephson junctions},\ }\href@noop {} {\bibfield  {journal}
  {\bibinfo  {journal} {arXiv:2307.04789}\ } (\bibinfo {year}
  {2023})}\BibitemShut {NoStop}%
\bibitem [{\citenamefont {Shen}\ \emph {et~al.}(2024)\citenamefont {Shen},
  \citenamefont {Lu}, \citenamefont {Lado},\ and\ \citenamefont
  {Trif}}]{shen2024nonhermitian}%
  \BibitemOpen
  \bibfield  {author} {\bibinfo {author} {\bibfnamefont {P.-X.}\ \bibnamefont
  {Shen}}, \bibinfo {author} {\bibfnamefont {Z.}~\bibnamefont {Lu}}, \bibinfo
  {author} {\bibfnamefont {J.~L.}\ \bibnamefont {Lado}},\ and\ \bibinfo
  {author} {\bibfnamefont {M.}~\bibnamefont {Trif}},\ }\bibfield  {title}
  {\bibinfo {title} {Non-{H}ermitian persistent current transport},\
  }\href@noop {} {\bibfield  {journal} {\bibinfo  {journal} {arXiv:2403.09569}\
  } (\bibinfo {year} {2024})}\BibitemShut {NoStop}%
\bibitem [{\citenamefont {Beenakker}(2024)}]{beenakker2024josephson}%
  \BibitemOpen
  \bibfield  {author} {\bibinfo {author} {\bibfnamefont {C.~W.~J.}\
  \bibnamefont {Beenakker}},\ }\bibfield  {title} {\bibinfo {title} {Josephson
  effect in a junction coupled to an electron reservoir},\ }\href@noop {}
  {\bibfield  {journal} {\bibinfo  {journal} {arXiv:2404.13976}\ } (\bibinfo
  {year} {2024})}\BibitemShut {NoStop}%
\bibitem [{\citenamefont {Pino}\ \emph {et~al.}(2024)\citenamefont {Pino},
  \citenamefont {Meir},\ and\ \citenamefont {Aguado}}]{pino2024thermodynamics}%
  \BibitemOpen
  \bibfield  {author} {\bibinfo {author} {\bibfnamefont {D.~M.}\ \bibnamefont
  {Pino}}, \bibinfo {author} {\bibfnamefont {Y.}~\bibnamefont {Meir}},\ and\
  \bibinfo {author} {\bibfnamefont {R.}~\bibnamefont {Aguado}},\ }\bibfield
  {title} {\bibinfo {title} {Thermodynamics of non-{H}ermitian {J}osephson
  junctions with exceptional points},\ }\href@noop {} {\bibfield  {journal}
  {\bibinfo  {journal} {arXiv:2405.02387}\ } (\bibinfo {year}
  {2024})}\BibitemShut {NoStop}%
\bibitem [{\citenamefont {Yokoyama}\ and\ \citenamefont
  {Nazarov}(2015)}]{PhysRevB.92.155437}%
  \BibitemOpen
  \bibfield  {author} {\bibinfo {author} {\bibfnamefont {T.}~\bibnamefont
  {Yokoyama}}\ and\ \bibinfo {author} {\bibfnamefont {Y.~V.}\ \bibnamefont
  {Nazarov}},\ }\bibfield  {title} {\bibinfo {title} {Singularities in the
  {A}ndreev spectrum of a multiterminal {J}osephson junction},\ }\href
  {https://doi.org/10.1103/PhysRevB.92.155437} {\bibfield  {journal} {\bibinfo
  {journal} {Phys. Rev. B}\ }\textbf {\bibinfo {volume} {92}},\ \bibinfo
  {pages} {155437} (\bibinfo {year} {2015})}\BibitemShut {NoStop}%
\bibitem [{\citenamefont {Riwar}\ \emph {et~al.}(2016)\citenamefont {Riwar},
  \citenamefont {Houzet}, \citenamefont {Meyer},\ and\ \citenamefont
  {Nazarov}}]{riwar2016multi}%
  \BibitemOpen
  \bibfield  {author} {\bibinfo {author} {\bibfnamefont {R.-P.}\ \bibnamefont
  {Riwar}}, \bibinfo {author} {\bibfnamefont {M.}~\bibnamefont {Houzet}},
  \bibinfo {author} {\bibfnamefont {J.~S.}\ \bibnamefont {Meyer}},\ and\
  \bibinfo {author} {\bibfnamefont {Y.~V.}\ \bibnamefont {Nazarov}},\
  }\bibfield  {title} {\bibinfo {title} {Multi-terminal {J}osephson junctions
  as topological matter},\ }\href@noop {} {\bibfield  {journal} {\bibinfo
  {journal} {Nat. Commun.}\ }\textbf {\bibinfo {volume} {7}},\ \bibinfo {pages}
  {11167} (\bibinfo {year} {2016})}\BibitemShut {NoStop}%
\bibitem [{\citenamefont {Peralta~Gavensky}\ \emph {et~al.}(2022)\citenamefont
  {Peralta~Gavensky}, \citenamefont {Usaj},\ and\ \citenamefont
  {Balseiro}}]{peralta2022multi}%
  \BibitemOpen
  \bibfield  {author} {\bibinfo {author} {\bibfnamefont {L.}~\bibnamefont
  {Peralta~Gavensky}}, \bibinfo {author} {\bibfnamefont {G.}~\bibnamefont
  {Usaj}},\ and\ \bibinfo {author} {\bibfnamefont {C.~A.}\ \bibnamefont
  {Balseiro}},\ }\bibfield  {title} {\bibinfo {title} {Multi-terminal
  {J}osephson junctions: {A} road to topological flux networks},\ }\href
  {http://dx.doi.org/10.1209/0295-5075/acb2f6} {\bibfield  {journal} {\bibinfo
  {journal} {Europhys. Lett.}\ }\textbf {\bibinfo {volume} {141}},\ \bibinfo
  {pages} {36001} (\bibinfo {year} {2022})}\BibitemShut {NoStop}%
\bibitem [{\citenamefont {Su}\ \emph {et~al.}(2017)\citenamefont {Su},
  \citenamefont {Tacla}, \citenamefont {Hocevar}, \citenamefont {Car},
  \citenamefont {Plissard}, \citenamefont {Bakkers}, \citenamefont {Daley},
  \citenamefont {Pekker},\ and\ \citenamefont {Frolov}}]{Su_2017}%
  \BibitemOpen
  \bibfield  {author} {\bibinfo {author} {\bibfnamefont {Z.}~\bibnamefont
  {Su}}, \bibinfo {author} {\bibfnamefont {A.~B.}\ \bibnamefont {Tacla}},
  \bibinfo {author} {\bibfnamefont {M.}~\bibnamefont {Hocevar}}, \bibinfo
  {author} {\bibfnamefont {D.}~\bibnamefont {Car}}, \bibinfo {author}
  {\bibfnamefont {S.~R.}\ \bibnamefont {Plissard}}, \bibinfo {author}
  {\bibfnamefont {E.~P. A.~M.}\ \bibnamefont {Bakkers}}, \bibinfo {author}
  {\bibfnamefont {A.~J.}\ \bibnamefont {Daley}}, \bibinfo {author}
  {\bibfnamefont {D.}~\bibnamefont {Pekker}},\ and\ \bibinfo {author}
  {\bibfnamefont {S.~M.}\ \bibnamefont {Frolov}},\ }\bibfield  {title}
  {\bibinfo {title} {Andreev molecules in semiconductor nanowire double quantum
  dots},\ }\href {http://dx.doi.org/10.1038/s41467-017-00665-7} {\bibfield
  {journal} {\bibinfo  {journal} {Nat. Commun.}\ }\textbf {\bibinfo {volume}
  {8}},\ \bibinfo {pages} {585} (\bibinfo {year} {2017})}\BibitemShut {NoStop}%
\bibitem [{\citenamefont {Pillet}\ \emph {et~al.}(2020)\citenamefont {Pillet},
  \citenamefont {Benzoni}, \citenamefont {Griesmar}, \citenamefont {Smirr},\
  and\ \citenamefont {\c{C}aglar
  \"{O}.~Girit}}]{10.21468/SciPostPhysCore.2.2.009}%
  \BibitemOpen
  \bibfield  {author} {\bibinfo {author} {\bibfnamefont {J.-D.}\ \bibnamefont
  {Pillet}}, \bibinfo {author} {\bibfnamefont {V.}~\bibnamefont {Benzoni}},
  \bibinfo {author} {\bibfnamefont {J.}~\bibnamefont {Griesmar}}, \bibinfo
  {author} {\bibfnamefont {J.-L.}\ \bibnamefont {Smirr}},\ and\ \bibinfo
  {author} {\bibnamefont {\c{C}aglar \"{O}.~Girit}},\ }\bibfield  {title}
  {\bibinfo {title} {{Scattering description of {A}ndreev molecules}},\ }\href
  {https://doi.org/10.21468/SciPostPhysCore.2.2.009} {\bibfield  {journal}
  {\bibinfo  {journal} {SciPost Phys. Core}\ }\textbf {\bibinfo {volume} {2}},\
  \bibinfo {pages} {009} (\bibinfo {year} {2020})}\BibitemShut {NoStop}%
\bibitem [{\citenamefont {Sakurai}\ \emph {et~al.}(2020)\citenamefont
  {Sakurai}, \citenamefont {Mercaldo}, \citenamefont {Kobayashi}, \citenamefont
  {Yamakage}, \citenamefont {Ikegaya}, \citenamefont {Habe}, \citenamefont
  {Kotetes}, \citenamefont {Cuoco},\ and\ \citenamefont
  {Asano}}]{PhysRevB.101.174506}%
  \BibitemOpen
  \bibfield  {author} {\bibinfo {author} {\bibfnamefont {K.}~\bibnamefont
  {Sakurai}}, \bibinfo {author} {\bibfnamefont {M.~T.}\ \bibnamefont
  {Mercaldo}}, \bibinfo {author} {\bibfnamefont {S.}~\bibnamefont {Kobayashi}},
  \bibinfo {author} {\bibfnamefont {A.}~\bibnamefont {Yamakage}}, \bibinfo
  {author} {\bibfnamefont {S.}~\bibnamefont {Ikegaya}}, \bibinfo {author}
  {\bibfnamefont {T.}~\bibnamefont {Habe}}, \bibinfo {author} {\bibfnamefont
  {P.}~\bibnamefont {Kotetes}}, \bibinfo {author} {\bibfnamefont
  {M.}~\bibnamefont {Cuoco}},\ and\ \bibinfo {author} {\bibfnamefont
  {Y.}~\bibnamefont {Asano}},\ }\bibfield  {title} {\bibinfo {title} {Nodal
  {A}ndreev spectra in multi-{M}ajorana three-terminal {J}osephson junctions},\
  }\href {https://doi.org/10.1103/PhysRevB.101.174506} {\bibfield  {journal}
  {\bibinfo  {journal} {Phys. Rev. B}\ }\textbf {\bibinfo {volume} {101}},\
  \bibinfo {pages} {174506} (\bibinfo {year} {2020})}\BibitemShut {NoStop}%
\bibitem [{\citenamefont {Matsuo}\ \emph
  {et~al.}(2023{\natexlab{a}})\citenamefont {Matsuo}, \citenamefont {Imoto},
  \citenamefont {Yokoyama}, \citenamefont {Sato}, \citenamefont {Lindemann},
  \citenamefont {Gronin}, \citenamefont {Gardner}, \citenamefont {Nakosai},
  \citenamefont {Tanaka}, \citenamefont {Manfra} \emph
  {et~al.}}]{matsuo2023phase2}%
  \BibitemOpen
  \bibfield  {author} {\bibinfo {author} {\bibfnamefont {S.}~\bibnamefont
  {Matsuo}}, \bibinfo {author} {\bibfnamefont {T.}~\bibnamefont {Imoto}},
  \bibinfo {author} {\bibfnamefont {T.}~\bibnamefont {Yokoyama}}, \bibinfo
  {author} {\bibfnamefont {Y.}~\bibnamefont {Sato}}, \bibinfo {author}
  {\bibfnamefont {T.}~\bibnamefont {Lindemann}}, \bibinfo {author}
  {\bibfnamefont {S.}~\bibnamefont {Gronin}}, \bibinfo {author} {\bibfnamefont
  {G.~C.}\ \bibnamefont {Gardner}}, \bibinfo {author} {\bibfnamefont
  {S.}~\bibnamefont {Nakosai}}, \bibinfo {author} {\bibfnamefont
  {Y.}~\bibnamefont {Tanaka}}, \bibinfo {author} {\bibfnamefont {M.~J.}\
  \bibnamefont {Manfra}}, \emph {et~al.},\ }\bibfield  {title} {\bibinfo
  {title} {Phase-dependent {A}ndreev molecules and superconducting gap closing
  in coherently-coupled {J}osephson junctions},\ }\href@noop {} {\bibfield
  {journal} {\bibinfo  {journal} {Nat. Commun.}\ }\textbf {\bibinfo {volume}
  {14}},\ \bibinfo {pages} {8271} (\bibinfo {year}
  {2023}{\natexlab{a}})}\BibitemShut {NoStop}%
\bibitem [{\citenamefont {Coraiola}\ \emph {et~al.}(2023)\citenamefont
  {Coraiola}, \citenamefont {Haxell}, \citenamefont {Sabonis}, \citenamefont
  {Weisbrich}, \citenamefont {Svetogorov}, \citenamefont {Hinderling},
  \citenamefont {ten Kate}, \citenamefont {Cheah}, \citenamefont {Krizek},
  \citenamefont {Schott}, \citenamefont {Wegscheider}, \citenamefont {Cuevas},
  \citenamefont {Belzig},\ and\ \citenamefont {Nichele}}]{Coraiola_2023}%
  \BibitemOpen
  \bibfield  {author} {\bibinfo {author} {\bibfnamefont {M.}~\bibnamefont
  {Coraiola}}, \bibinfo {author} {\bibfnamefont {D.~Z.}\ \bibnamefont
  {Haxell}}, \bibinfo {author} {\bibfnamefont {D.}~\bibnamefont {Sabonis}},
  \bibinfo {author} {\bibfnamefont {H.}~\bibnamefont {Weisbrich}}, \bibinfo
  {author} {\bibfnamefont {A.~E.}\ \bibnamefont {Svetogorov}}, \bibinfo
  {author} {\bibfnamefont {M.}~\bibnamefont {Hinderling}}, \bibinfo {author}
  {\bibfnamefont {S.~C.}\ \bibnamefont {ten Kate}}, \bibinfo {author}
  {\bibfnamefont {E.}~\bibnamefont {Cheah}}, \bibinfo {author} {\bibfnamefont
  {F.}~\bibnamefont {Krizek}}, \bibinfo {author} {\bibfnamefont
  {R.}~\bibnamefont {Schott}}, \bibinfo {author} {\bibfnamefont
  {W.}~\bibnamefont {Wegscheider}}, \bibinfo {author} {\bibfnamefont {J.~C.}\
  \bibnamefont {Cuevas}}, \bibinfo {author} {\bibfnamefont {W.}~\bibnamefont
  {Belzig}},\ and\ \bibinfo {author} {\bibfnamefont {F.}~\bibnamefont
  {Nichele}},\ }\bibfield  {title} {\bibinfo {title} {Phase-engineering the
  {A}ndreev band structure of a three-terminal {J}osephson junction},\ }\href
  {http://dx.doi.org/10.1038/s41467-023-42356-6} {\bibfield  {journal}
  {\bibinfo  {journal} {Nat. Commun.}\ }\textbf {\bibinfo {volume} {14}},\
  \bibinfo {pages} {6784} (\bibinfo {year} {2023})}\BibitemShut {NoStop}%
\bibitem [{\citenamefont {Matute-Ca\~nadas}\ \emph {et~al.}(2024)\citenamefont
  {Matute-Ca\~nadas}, \citenamefont {Tosi},\ and\ \citenamefont
  {Yeyati}}]{PRXQuantum.5.020340}%
  \BibitemOpen
  \bibfield  {author} {\bibinfo {author} {\bibfnamefont {F.}~\bibnamefont
  {Matute-Ca\~nadas}}, \bibinfo {author} {\bibfnamefont {L.}~\bibnamefont
  {Tosi}},\ and\ \bibinfo {author} {\bibfnamefont {A.~L.}\ \bibnamefont
  {Yeyati}},\ }\bibfield  {title} {\bibinfo {title} {Quantum circuits with
  multiterminal {J}osephson-{A}ndreev junctions},\ }\href
  {https://doi.org/10.1103/PRXQuantum.5.020340} {\bibfield  {journal} {\bibinfo
   {journal} {PRX Quantum}\ }\textbf {\bibinfo {volume} {5}},\ \bibinfo {pages}
  {020340} (\bibinfo {year} {2024})}\BibitemShut {NoStop}%
\bibitem [{\citenamefont {Pillet}\ \emph {et~al.}(2019)\citenamefont {Pillet},
  \citenamefont {Benzoni}, \citenamefont {Griesmar}, \citenamefont {Smirr},\
  and\ \citenamefont {Girit}}]{Pillet_2019}%
  \BibitemOpen
  \bibfield  {author} {\bibinfo {author} {\bibfnamefont {J.-D.}\ \bibnamefont
  {Pillet}}, \bibinfo {author} {\bibfnamefont {V.}~\bibnamefont {Benzoni}},
  \bibinfo {author} {\bibfnamefont {J.}~\bibnamefont {Griesmar}}, \bibinfo
  {author} {\bibfnamefont {J.-L.}\ \bibnamefont {Smirr}},\ and\ \bibinfo
  {author} {\bibfnamefont {c.~O.}\ \bibnamefont {Girit}},\ }\bibfield  {title}
  {\bibinfo {title} {Nonlocal {J}osephson effect in {A}ndreev molecules},\
  }\href {http://dx.doi.org/10.1021/acs.nanolett.9b02686} {\bibfield  {journal}
  {\bibinfo  {journal} {Nano Letters}\ }\textbf {\bibinfo {volume} {19}},\
  \bibinfo {pages} {7138} (\bibinfo {year} {2019})}\BibitemShut {NoStop}%
\bibitem [{\citenamefont {Pankratova}\ \emph {et~al.}(2020)\citenamefont
  {Pankratova}, \citenamefont {Lee}, \citenamefont {Kuzmin}, \citenamefont
  {Wickramasinghe}, \citenamefont {Mayer}, \citenamefont {Yuan}, \citenamefont
  {Vavilov}, \citenamefont {Shabani},\ and\ \citenamefont
  {Manucharyan}}]{PhysRevX.10.031051}%
  \BibitemOpen
  \bibfield  {author} {\bibinfo {author} {\bibfnamefont {N.}~\bibnamefont
  {Pankratova}}, \bibinfo {author} {\bibfnamefont {H.}~\bibnamefont {Lee}},
  \bibinfo {author} {\bibfnamefont {R.}~\bibnamefont {Kuzmin}}, \bibinfo
  {author} {\bibfnamefont {K.}~\bibnamefont {Wickramasinghe}}, \bibinfo
  {author} {\bibfnamefont {W.}~\bibnamefont {Mayer}}, \bibinfo {author}
  {\bibfnamefont {J.}~\bibnamefont {Yuan}}, \bibinfo {author} {\bibfnamefont
  {M.~G.}\ \bibnamefont {Vavilov}}, \bibinfo {author} {\bibfnamefont
  {J.}~\bibnamefont {Shabani}},\ and\ \bibinfo {author} {\bibfnamefont {V.~E.}\
  \bibnamefont {Manucharyan}},\ }\bibfield  {title} {\bibinfo {title}
  {Multiterminal josephson effect},\ }\href
  {https://doi.org/10.1103/PhysRevX.10.031051} {\bibfield  {journal} {\bibinfo
  {journal} {Phys. Rev. X}\ }\textbf {\bibinfo {volume} {10}},\ \bibinfo
  {pages} {031051} (\bibinfo {year} {2020})}\BibitemShut {NoStop}%
\bibitem [{\citenamefont {Matsuo}\ \emph {et~al.}(2022)\citenamefont {Matsuo},
  \citenamefont {Lee}, \citenamefont {Chang}, \citenamefont {Sato},
  \citenamefont {Ueda}, \citenamefont {Palmstr{\o}m},\ and\ \citenamefont
  {Tarucha}}]{Matsuo_2022}%
  \BibitemOpen
  \bibfield  {author} {\bibinfo {author} {\bibfnamefont {S.}~\bibnamefont
  {Matsuo}}, \bibinfo {author} {\bibfnamefont {J.~S.}\ \bibnamefont {Lee}},
  \bibinfo {author} {\bibfnamefont {C.-Y.}\ \bibnamefont {Chang}}, \bibinfo
  {author} {\bibfnamefont {Y.}~\bibnamefont {Sato}}, \bibinfo {author}
  {\bibfnamefont {K.}~\bibnamefont {Ueda}}, \bibinfo {author} {\bibfnamefont
  {C.~J.}\ \bibnamefont {Palmstr{\o}m}},\ and\ \bibinfo {author} {\bibfnamefont
  {S.}~\bibnamefont {Tarucha}},\ }\bibfield  {title} {\bibinfo {title}
  {Observation of nonlocal josephson effect on double inas nanowires},\ }\href
  {http://dx.doi.org/10.1038/s42005-022-00994-0} {\bibfield  {journal}
  {\bibinfo  {journal} {Commun. Phys.}\ }\textbf {\bibinfo {volume} {5}},\
  \bibinfo {pages} {221} (\bibinfo {year} {2022})}\BibitemShut {NoStop}%
\bibitem [{\citenamefont {Matsuo}\ \emph
  {et~al.}(2023{\natexlab{b}})\citenamefont {Matsuo}, \citenamefont {Imoto},
  \citenamefont {Yokoyama}, \citenamefont {Sato}, \citenamefont {Lindemann},
  \citenamefont {Gronin}, \citenamefont {Gardner}, \citenamefont {Manfra},\
  and\ \citenamefont {Tarucha}}]{matsuo2023phase}%
  \BibitemOpen
  \bibfield  {author} {\bibinfo {author} {\bibfnamefont {S.}~\bibnamefont
  {Matsuo}}, \bibinfo {author} {\bibfnamefont {T.}~\bibnamefont {Imoto}},
  \bibinfo {author} {\bibfnamefont {T.}~\bibnamefont {Yokoyama}}, \bibinfo
  {author} {\bibfnamefont {Y.}~\bibnamefont {Sato}}, \bibinfo {author}
  {\bibfnamefont {T.}~\bibnamefont {Lindemann}}, \bibinfo {author}
  {\bibfnamefont {S.}~\bibnamefont {Gronin}}, \bibinfo {author} {\bibfnamefont
  {G.~C.}\ \bibnamefont {Gardner}}, \bibinfo {author} {\bibfnamefont {M.~J.}\
  \bibnamefont {Manfra}},\ and\ \bibinfo {author} {\bibfnamefont
  {S.}~\bibnamefont {Tarucha}},\ }\bibfield  {title} {\bibinfo {title} {Phase
  engineering of anomalous {J}osephson effect derived from {A}ndreev
  molecules},\ }\href@noop {} {\bibfield  {journal} {\bibinfo  {journal}
  {Science Advances}\ }\textbf {\bibinfo {volume} {9}},\ \bibinfo {pages}
  {eadj3698} (\bibinfo {year} {2023}{\natexlab{b}})}\BibitemShut {NoStop}%
\bibitem [{\citenamefont {Haxell}\ \emph {et~al.}(2023)\citenamefont {Haxell},
  \citenamefont {Coraiola}, \citenamefont {Hinderling}, \citenamefont {ten
  Kate}, \citenamefont {Sabonis}, \citenamefont {Svetogorov}, \citenamefont
  {Belzig}, \citenamefont {Cheah}, \citenamefont {Krizek}, \citenamefont
  {Schott}, \citenamefont {Wegscheider},\ and\ \citenamefont
  {Nichele}}]{Haxell_2023}%
  \BibitemOpen
  \bibfield  {author} {\bibinfo {author} {\bibfnamefont {D.~Z.}\ \bibnamefont
  {Haxell}}, \bibinfo {author} {\bibfnamefont {M.}~\bibnamefont {Coraiola}},
  \bibinfo {author} {\bibfnamefont {M.}~\bibnamefont {Hinderling}}, \bibinfo
  {author} {\bibfnamefont {S.~C.}\ \bibnamefont {ten Kate}}, \bibinfo {author}
  {\bibfnamefont {D.}~\bibnamefont {Sabonis}}, \bibinfo {author} {\bibfnamefont
  {A.~E.}\ \bibnamefont {Svetogorov}}, \bibinfo {author} {\bibfnamefont
  {W.}~\bibnamefont {Belzig}}, \bibinfo {author} {\bibfnamefont
  {E.}~\bibnamefont {Cheah}}, \bibinfo {author} {\bibfnamefont
  {F.}~\bibnamefont {Krizek}}, \bibinfo {author} {\bibfnamefont
  {R.}~\bibnamefont {Schott}}, \bibinfo {author} {\bibfnamefont
  {W.}~\bibnamefont {Wegscheider}},\ and\ \bibinfo {author} {\bibfnamefont
  {F.}~\bibnamefont {Nichele}},\ }\bibfield  {title} {\bibinfo {title}
  {Demonstration of the nonlocal {J}osephson effect in {A}ndreev molecules},\
  }\href {http://dx.doi.org/10.1021/acs.nanolett.3c02066} {\bibfield  {journal}
  {\bibinfo  {journal} {Nano Letters}\ }\textbf {\bibinfo {volume} {23}},\
  \bibinfo {pages} {7532} (\bibinfo {year} {2023})}\BibitemShut {NoStop}%
\bibitem [{\citenamefont {Cayao}\ \emph {et~al.}(2024)\citenamefont {Cayao},
  \citenamefont {Burset},\ and\ \citenamefont {Tanaka}}]{PhysRevB.109.205406}%
  \BibitemOpen
  \bibfield  {author} {\bibinfo {author} {\bibfnamefont {J.}~\bibnamefont
  {Cayao}}, \bibinfo {author} {\bibfnamefont {P.}~\bibnamefont {Burset}},\ and\
  \bibinfo {author} {\bibfnamefont {Y.}~\bibnamefont {Tanaka}},\ }\bibfield
  {title} {\bibinfo {title} {Controllable odd-frequency {C}ooper pairs in
  multisuperconductor {J}osephson junctions},\ }\href
  {https://doi.org/10.1103/PhysRevB.109.205406} {\bibfield  {journal} {\bibinfo
   {journal} {Phys. Rev. B}\ }\textbf {\bibinfo {volume} {109}},\ \bibinfo
  {pages} {205406} (\bibinfo {year} {2024})}\BibitemShut {NoStop}%
\bibitem [{\citenamefont {Correa}\ and\ \citenamefont
  {Nowak}(2024)}]{10.21468/SciPostPhys.17.2.037}%
  \BibitemOpen
  \bibfield  {author} {\bibinfo {author} {\bibfnamefont {J.~H.}\ \bibnamefont
  {Correa}}\ and\ \bibinfo {author} {\bibfnamefont {M.~P.}\ \bibnamefont
  {Nowak}},\ }\bibfield  {title} {\bibinfo {title} {{Theory of universal diode
  effect in three-terminal Josephson junctions}},\ }\href
  {https://doi.org/10.21468/SciPostPhys.17.2.037} {\bibfield  {journal}
  {\bibinfo  {journal} {SciPost Phys.}\ }\textbf {\bibinfo {volume} {17}},\
  \bibinfo {pages} {037} (\bibinfo {year} {2024})}\BibitemShut {NoStop}%
\bibitem [{\citenamefont {Tinkham}(2004)}]{Tinkham}%
  \BibitemOpen
  \bibfield  {author} {\bibinfo {author} {\bibfnamefont {M.}~\bibnamefont
  {Tinkham}},\ }\href@noop {} {\emph {\bibinfo {title} {Introduction to
  superconductivity}}}\ (\bibinfo  {publisher} {Courier Corporation},\ \bibinfo
  {year} {2004})\BibitemShut {NoStop}%
\bibitem [{\citenamefont {Leijnse}\ and\ \citenamefont
  {Flensberg}(2012)}]{PhysRevB.86.134528}%
  \BibitemOpen
  \bibfield  {author} {\bibinfo {author} {\bibfnamefont {M.}~\bibnamefont
  {Leijnse}}\ and\ \bibinfo {author} {\bibfnamefont {K.}~\bibnamefont
  {Flensberg}},\ }\bibfield  {title} {\bibinfo {title} {Parity qubits and poor
  man's {M}ajorana bound states in double quantum dots},\ }\href
  {https://doi.org/10.1103/PhysRevB.86.134528} {\bibfield  {journal} {\bibinfo
  {journal} {Phys. Rev. B}\ }\textbf {\bibinfo {volume} {86}},\ \bibinfo
  {pages} {134528} (\bibinfo {year} {2012})}\BibitemShut {NoStop}%
\bibitem [{\citenamefont {Cayao}\ and\ \citenamefont
  {Aguado}(2024)}]{cayao2024NHtwositeKitaev}%
  \BibitemOpen
  \bibfield  {author} {\bibinfo {author} {\bibfnamefont {J.}~\bibnamefont
  {Cayao}}\ and\ \bibinfo {author} {\bibfnamefont {R.}~\bibnamefont {Aguado}},\
  }\bibfield  {title} {\bibinfo {title} {Non-{H}ermitian minimal {K}itaev
  chains},\ }\href {https://arxiv.org/abs/2406.18974} {\bibfield  {journal}
  {\bibinfo  {journal} {arXiv:2406.18974}\ } (\bibinfo {year}
  {2024})}\BibitemShut {NoStop}%
\bibitem [{\citenamefont {Cayao}(2024{\natexlab{b}})}]{cayao2024pairPMMMs}%
  \BibitemOpen
  \bibfield  {author} {\bibinfo {author} {\bibfnamefont {J.}~\bibnamefont
  {Cayao}},\ }\bibfield  {title} {\bibinfo {title} {Emergent pair symmetries in
  systems with poor man's {M}ajorana modes},\ }\href
  {https://doi.org/10.1103/PhysRevB.110.125408} {\bibfield  {journal} {\bibinfo
   {journal} {Phys. Rev. B}\ }\textbf {\bibinfo {volume} {110}},\ \bibinfo
  {pages} {125408} (\bibinfo {year} {2024}{\natexlab{b}})}\BibitemShut
  {NoStop}%
\bibitem [{\citenamefont {Tanaka}\ \emph {et~al.}(2024)\citenamefont {Tanaka},
  \citenamefont {Tamura},\ and\ \citenamefont {Cayao}}]{tanaka2024theory}%
  \BibitemOpen
  \bibfield  {author} {\bibinfo {author} {\bibfnamefont {Y.}~\bibnamefont
  {Tanaka}}, \bibinfo {author} {\bibfnamefont {S.}~\bibnamefont {Tamura}},\
  and\ \bibinfo {author} {\bibfnamefont {J.}~\bibnamefont {Cayao}},\ }\bibfield
   {title} {\bibinfo {title} {Theory of {M}ajorana zero modes in unconventional
  superconductors},\ }\href {https://doi.org/10.1093/ptep/ptae065} {\bibfield
  {journal} {\bibinfo  {journal} {Prog. Theor. Exp. Phys.}\ ,\ \bibinfo {pages}
  {ptae065}} (\bibinfo {year} {2024})}\BibitemShut {NoStop}%
\bibitem [{\citenamefont {Kawabata}\ \emph
  {et~al.}(2019{\natexlab{b}})\citenamefont {Kawabata}, \citenamefont
  {Bessho},\ and\ \citenamefont {Sato}}]{KBS19}%
  \BibitemOpen
  \bibfield  {author} {\bibinfo {author} {\bibfnamefont {K.}~\bibnamefont
  {Kawabata}}, \bibinfo {author} {\bibfnamefont {T.}~\bibnamefont {Bessho}},\
  and\ \bibinfo {author} {\bibfnamefont {M.}~\bibnamefont {Sato}},\ }\bibfield
  {title} {\bibinfo {title} {Classification of exceptional points and
  non-{H}ermitian topological semimetals},\ }\href
  {https://doi.org/10.1103/PhysRevLett.123.066405} {\bibfield  {journal}
  {\bibinfo  {journal} {Phys. Rev. Lett.}\ }\textbf {\bibinfo {volume} {123}},\
  \bibinfo {pages} {066405} (\bibinfo {year} {2019}{\natexlab{b}})}\BibitemShut
  {NoStop}%
\bibitem [{\citenamefont {Bergeret}\ \emph {et~al.}(2005)\citenamefont
  {Bergeret}, \citenamefont {Volkov},\ and\ \citenamefont
  {Efetov}}]{RevModPhys.77.1321}%
  \BibitemOpen
  \bibfield  {author} {\bibinfo {author} {\bibfnamefont {F.~S.}\ \bibnamefont
  {Bergeret}}, \bibinfo {author} {\bibfnamefont {A.~F.}\ \bibnamefont
  {Volkov}},\ and\ \bibinfo {author} {\bibfnamefont {K.~B.}\ \bibnamefont
  {Efetov}},\ }\bibfield  {title} {\bibinfo {title} {Odd triplet
  superconductivity and related phenomena in superconductor-ferromagnet
  structures},\ }\href {https://doi.org/10.1103/RevModPhys.77.1321} {\bibfield
  {journal} {\bibinfo  {journal} {Rev. Mod. Phys.}\ }\textbf {\bibinfo {volume}
  {77}},\ \bibinfo {pages} {1321} (\bibinfo {year} {2005})}\BibitemShut
  {NoStop}%
\bibitem [{\citenamefont {Dvir}\ \emph {et~al.}(2023)\citenamefont {Dvir},
  \citenamefont {Wang}, \citenamefont {van Loo}, \citenamefont {Liu},
  \citenamefont {Mazur}, \citenamefont {Bordin}, \citenamefont {Ten~Haaf},
  \citenamefont {Wang}, \citenamefont {van Driel}, \citenamefont {Zatelli}
  \emph {et~al.}}]{dvir2023realization}%
  \BibitemOpen
  \bibfield  {author} {\bibinfo {author} {\bibfnamefont {T.}~\bibnamefont
  {Dvir}}, \bibinfo {author} {\bibfnamefont {G.}~\bibnamefont {Wang}}, \bibinfo
  {author} {\bibfnamefont {N.}~\bibnamefont {van Loo}}, \bibinfo {author}
  {\bibfnamefont {C.-X.}\ \bibnamefont {Liu}}, \bibinfo {author} {\bibfnamefont
  {G.~P.}\ \bibnamefont {Mazur}}, \bibinfo {author} {\bibfnamefont
  {A.}~\bibnamefont {Bordin}}, \bibinfo {author} {\bibfnamefont {S.~L.}\
  \bibnamefont {Ten~Haaf}}, \bibinfo {author} {\bibfnamefont {J.-Y.}\
  \bibnamefont {Wang}}, \bibinfo {author} {\bibfnamefont {D.}~\bibnamefont {van
  Driel}}, \bibinfo {author} {\bibfnamefont {F.}~\bibnamefont {Zatelli}}, \emph
  {et~al.},\ }\bibfield  {title} {\bibinfo {title} {Realization of a minimal
  {K}itaev chain in coupled quantum dots},\ }\href@noop {} {\bibfield
  {journal} {\bibinfo  {journal} {Nature}\ }\textbf {\bibinfo {volume} {614}},\
  \bibinfo {pages} {445} (\bibinfo {year} {2023})}\BibitemShut {NoStop}%
\bibitem [{\citenamefont {Bordin}\ \emph
  {et~al.}(2023{\natexlab{a}})\citenamefont {Bordin}, \citenamefont {Li},
  \citenamefont {van Driel}, \citenamefont {Wolff}, \citenamefont {Wang},
  \citenamefont {ten Haaf}, \citenamefont {Wang}, \citenamefont {van Loo},
  \citenamefont {Kouwenhoven},\ and\ \citenamefont {Dvir}}]{bordin2023crossed}%
  \BibitemOpen
  \bibfield  {author} {\bibinfo {author} {\bibfnamefont {A.}~\bibnamefont
  {Bordin}}, \bibinfo {author} {\bibfnamefont {X.}~\bibnamefont {Li}}, \bibinfo
  {author} {\bibfnamefont {D.}~\bibnamefont {van Driel}}, \bibinfo {author}
  {\bibfnamefont {J.~C.}\ \bibnamefont {Wolff}}, \bibinfo {author}
  {\bibfnamefont {Q.}~\bibnamefont {Wang}}, \bibinfo {author} {\bibfnamefont
  {S.~L.~D.}\ \bibnamefont {ten Haaf}}, \bibinfo {author} {\bibfnamefont
  {G.}~\bibnamefont {Wang}}, \bibinfo {author} {\bibfnamefont {N.}~\bibnamefont
  {van Loo}}, \bibinfo {author} {\bibfnamefont {L.~P.}\ \bibnamefont
  {Kouwenhoven}},\ and\ \bibinfo {author} {\bibfnamefont {T.}~\bibnamefont
  {Dvir}},\ }\bibfield  {title} {\bibinfo {title} {Crossed {A}ndreev reflection
  and elastic co-tunneling in a three-site kitaev chain nanowire device},\
  }\href@noop {} {\bibfield  {journal} {\bibinfo  {journal} {arXiv:2306.07696}\
  } (\bibinfo {year} {2023}{\natexlab{a}})}\BibitemShut {NoStop}%
\bibitem [{\citenamefont {Bordin}\ \emph
  {et~al.}(2023{\natexlab{b}})\citenamefont {Bordin}, \citenamefont {Wang},
  \citenamefont {Liu}, \citenamefont {ten Haaf}, \citenamefont {van Loo},
  \citenamefont {Mazur}, \citenamefont {Xu}, \citenamefont {van Driel},
  \citenamefont {Zatelli}, \citenamefont {Gazibegovic}, \citenamefont {Badawy},
  \citenamefont {Bakkers}, \citenamefont {Wimmer}, \citenamefont
  {Kouwenhoven},\ and\ \citenamefont {Dvir}}]{PhysRevX.13.031031}%
  \BibitemOpen
  \bibfield  {author} {\bibinfo {author} {\bibfnamefont {A.}~\bibnamefont
  {Bordin}}, \bibinfo {author} {\bibfnamefont {G.}~\bibnamefont {Wang}},
  \bibinfo {author} {\bibfnamefont {C.-X.}\ \bibnamefont {Liu}}, \bibinfo
  {author} {\bibfnamefont {S.~L.~D.}\ \bibnamefont {ten Haaf}}, \bibinfo
  {author} {\bibfnamefont {N.}~\bibnamefont {van Loo}}, \bibinfo {author}
  {\bibfnamefont {G.~P.}\ \bibnamefont {Mazur}}, \bibinfo {author}
  {\bibfnamefont {D.}~\bibnamefont {Xu}}, \bibinfo {author} {\bibfnamefont
  {D.}~\bibnamefont {van Driel}}, \bibinfo {author} {\bibfnamefont
  {F.}~\bibnamefont {Zatelli}}, \bibinfo {author} {\bibfnamefont
  {S.}~\bibnamefont {Gazibegovic}}, \bibinfo {author} {\bibfnamefont
  {G.}~\bibnamefont {Badawy}}, \bibinfo {author} {\bibfnamefont {E.~P. A.~M.}\
  \bibnamefont {Bakkers}}, \bibinfo {author} {\bibfnamefont {M.}~\bibnamefont
  {Wimmer}}, \bibinfo {author} {\bibfnamefont {L.~P.}\ \bibnamefont
  {Kouwenhoven}},\ and\ \bibinfo {author} {\bibfnamefont {T.}~\bibnamefont
  {Dvir}},\ }\bibfield  {title} {\bibinfo {title} {Tunable crossed {A}ndreev
  reflection and elastic cotunneling in hybrid nanowires},\ }\href
  {https://doi.org/10.1103/PhysRevX.13.031031} {\bibfield  {journal} {\bibinfo
  {journal} {Phys. Rev. X}\ }\textbf {\bibinfo {volume} {13}},\ \bibinfo
  {pages} {031031} (\bibinfo {year} {2023}{\natexlab{b}})}\BibitemShut
  {NoStop}%
\bibitem [{\citenamefont {Zatelli}\ \emph {et~al.}(2023)\citenamefont
  {Zatelli}, \citenamefont {van Driel}, \citenamefont {Xu}, \citenamefont
  {Wang}, \citenamefont {Liu}, \citenamefont {Bordin}, \citenamefont {Roovers},
  \citenamefont {Mazur}, \citenamefont {van Loo}, \citenamefont {Wolff},
  \citenamefont {Bozkurt}, \citenamefont {Badawy}, \citenamefont {Gazibegovic},
  \citenamefont {Bakkers}, \citenamefont {Wimmer}, \citenamefont
  {Kouwenhoven},\ and\ \citenamefont {Dvir}}]{zatelli2023robustpoorMajo}%
  \BibitemOpen
  \bibfield  {author} {\bibinfo {author} {\bibfnamefont {F.}~\bibnamefont
  {Zatelli}}, \bibinfo {author} {\bibfnamefont {D.}~\bibnamefont {van Driel}},
  \bibinfo {author} {\bibfnamefont {D.}~\bibnamefont {Xu}}, \bibinfo {author}
  {\bibfnamefont {G.}~\bibnamefont {Wang}}, \bibinfo {author} {\bibfnamefont
  {C.-X.}\ \bibnamefont {Liu}}, \bibinfo {author} {\bibfnamefont
  {A.}~\bibnamefont {Bordin}}, \bibinfo {author} {\bibfnamefont
  {B.}~\bibnamefont {Roovers}}, \bibinfo {author} {\bibfnamefont {G.~P.}\
  \bibnamefont {Mazur}}, \bibinfo {author} {\bibfnamefont {N.}~\bibnamefont
  {van Loo}}, \bibinfo {author} {\bibfnamefont {J.~C.}\ \bibnamefont {Wolff}},
  \bibinfo {author} {\bibfnamefont {A.~M.}\ \bibnamefont {Bozkurt}}, \bibinfo
  {author} {\bibfnamefont {G.}~\bibnamefont {Badawy}}, \bibinfo {author}
  {\bibfnamefont {S.}~\bibnamefont {Gazibegovic}}, \bibinfo {author}
  {\bibfnamefont {E.~P. A.~M.}\ \bibnamefont {Bakkers}}, \bibinfo {author}
  {\bibfnamefont {M.}~\bibnamefont {Wimmer}}, \bibinfo {author} {\bibfnamefont
  {L.~P.}\ \bibnamefont {Kouwenhoven}},\ and\ \bibinfo {author} {\bibfnamefont
  {T.}~\bibnamefont {Dvir}},\ }\bibfield  {title} {\bibinfo {title} {Robust
  poor man's {M}ajorana zero modes using {Y}u-{S}hiba-{R}usinov states},\
  }\href {https://arxiv.org/abs/2311.03193} {\bibfield  {journal} {\bibinfo
  {journal} {arXiv:2311.03193}\ } (\bibinfo {year} {2023})}\BibitemShut
  {NoStop}%
\bibitem [{\citenamefont {Strambini}\ \emph {et~al.}(2016)\citenamefont
  {Strambini}, \citenamefont {D'ambrosio}, \citenamefont {Vischi},
  \citenamefont {Bergeret}, \citenamefont {Nazarov},\ and\ \citenamefont
  {Giazotto}}]{strambini2016omega}%
  \BibitemOpen
  \bibfield  {author} {\bibinfo {author} {\bibfnamefont {E.}~\bibnamefont
  {Strambini}}, \bibinfo {author} {\bibfnamefont {S.}~\bibnamefont
  {D'ambrosio}}, \bibinfo {author} {\bibfnamefont {F.}~\bibnamefont {Vischi}},
  \bibinfo {author} {\bibfnamefont {F.}~\bibnamefont {Bergeret}}, \bibinfo
  {author} {\bibfnamefont {Y.~V.}\ \bibnamefont {Nazarov}},\ and\ \bibinfo
  {author} {\bibfnamefont {F.}~\bibnamefont {Giazotto}},\ }\bibfield  {title}
  {\bibinfo {title} {The $\omega$-squipt as a tool to phase-engineer
  {J}osephson topological materials},\ }\href@noop {} {\bibfield  {journal}
  {\bibinfo  {journal} {Nat. Nanotech.}\ }\textbf {\bibinfo {volume} {11}},\
  \bibinfo {pages} {1055} (\bibinfo {year} {2016})}\BibitemShut {NoStop}%
\bibitem [{\citenamefont {Draelos}\ \emph {et~al.}(2019)\citenamefont
  {Draelos}, \citenamefont {Wei}, \citenamefont {Seredinski}, \citenamefont
  {Li}, \citenamefont {Mehta}, \citenamefont {Watanabe}, \citenamefont
  {Taniguchi}, \citenamefont {Borzenets}, \citenamefont {Amet},\ and\
  \citenamefont {Finkelstein}}]{draelos2019supercurrent}%
  \BibitemOpen
  \bibfield  {author} {\bibinfo {author} {\bibfnamefont {A.~W.}\ \bibnamefont
  {Draelos}}, \bibinfo {author} {\bibfnamefont {M.-T.}\ \bibnamefont {Wei}},
  \bibinfo {author} {\bibfnamefont {A.}~\bibnamefont {Seredinski}}, \bibinfo
  {author} {\bibfnamefont {H.}~\bibnamefont {Li}}, \bibinfo {author}
  {\bibfnamefont {Y.}~\bibnamefont {Mehta}}, \bibinfo {author} {\bibfnamefont
  {K.}~\bibnamefont {Watanabe}}, \bibinfo {author} {\bibfnamefont
  {T.}~\bibnamefont {Taniguchi}}, \bibinfo {author} {\bibfnamefont {I.~V.}\
  \bibnamefont {Borzenets}}, \bibinfo {author} {\bibfnamefont {F.}~\bibnamefont
  {Amet}},\ and\ \bibinfo {author} {\bibfnamefont {G.}~\bibnamefont
  {Finkelstein}},\ }\bibfield  {title} {\bibinfo {title} {Supercurrent flow in
  multiterminal graphene {J}osephson junctions},\ }\href@noop {} {\bibfield
  {journal} {\bibinfo  {journal} {Nano Lett.}\ }\textbf {\bibinfo {volume}
  {19}},\ \bibinfo {pages} {1039} (\bibinfo {year} {2019})}\BibitemShut
  {NoStop}%
\bibitem [{\citenamefont {Graziano}\ \emph {et~al.}(2022)\citenamefont
  {Graziano}, \citenamefont {Gupta}, \citenamefont {Pendharkar}, \citenamefont
  {Dong}, \citenamefont {Dempsey}, \citenamefont {Palmstr{\o}m},\ and\
  \citenamefont {Pribiag}}]{graziano2022selective}%
  \BibitemOpen
  \bibfield  {author} {\bibinfo {author} {\bibfnamefont {G.~V.}\ \bibnamefont
  {Graziano}}, \bibinfo {author} {\bibfnamefont {M.}~\bibnamefont {Gupta}},
  \bibinfo {author} {\bibfnamefont {M.}~\bibnamefont {Pendharkar}}, \bibinfo
  {author} {\bibfnamefont {J.~T.}\ \bibnamefont {Dong}}, \bibinfo {author}
  {\bibfnamefont {C.~P.}\ \bibnamefont {Dempsey}}, \bibinfo {author}
  {\bibfnamefont {C.}~\bibnamefont {Palmstr{\o}m}},\ and\ \bibinfo {author}
  {\bibfnamefont {V.~S.}\ \bibnamefont {Pribiag}},\ }\bibfield  {title}
  {\bibinfo {title} {Selective control of conductance modes in multi-terminal
  {J}osephson junctions},\ }\href@noop {} {\bibfield  {journal} {\bibinfo
  {journal} {Nat. Commun.}\ }\textbf {\bibinfo {volume} {13}},\ \bibinfo
  {pages} {5933} (\bibinfo {year} {2022})}\BibitemShut {NoStop}%
\bibitem [{\citenamefont {Ohnmacht}\ \emph {et~al.}(2024)\citenamefont
  {Ohnmacht}, \citenamefont {Wilhelm}, \citenamefont {Weisbrich},\ and\
  \citenamefont {Belzig}}]{ohnmacht2024NHtopology}%
  \BibitemOpen
  \bibfield  {author} {\bibinfo {author} {\bibfnamefont {D.~C.}\ \bibnamefont
  {Ohnmacht}}, \bibinfo {author} {\bibfnamefont {V.}~\bibnamefont {Wilhelm}},
  \bibinfo {author} {\bibfnamefont {H.}~\bibnamefont {Weisbrich}},\ and\
  \bibinfo {author} {\bibfnamefont {W.}~\bibnamefont {Belzig}},\ }\bibfield
  {title} {\bibinfo {title} {Non-hermitian topology in multiterminal
  superconducting junctions},\ }\href {https://arxiv.org/abs/2408.01289}
  {\bibfield  {journal} {\bibinfo  {journal} {arXiv:2408.01289}\ } (\bibinfo
  {year} {2024})}\BibitemShut {NoStop}%
\end{thebibliography}%
\end{document}